# A convolutional autoencoder approach for mining features in cellular electron cryo-tomograms and weakly supervised coarse segmentation

Xiangrui Zeng[1], Miguel Ricardo Leung[2,3], Tzviya Zeev-Ben-Mordehai[2,3], and Min Xu*[1]

[1]Computational Biology Department, School of Computer Science, Carnegie Mellon University, Pittsburgh, 15213, USA
[2]Division of Structural Biology, Wellcome Trust Centre for Human Genetics, University of Oxford, Oxford, OX3 7BN, UK
[3]Cryo-electron Microscopy, Bijvoet Center for Biomolecular Research, Utrecht University, Utrecht, Netherlands

**Abstract**

Cellular electron cryo-tomography enables the 3D visualization of cellular organization in the near-native state and at submolecular resolution. However, the contents of cellular tomograms are often complex, making it difficult to automatically isolate different *in situ* cellular components. In this paper, we propose a convolutional autoencoder-based unsupervised approach to provide a coarse grouping of 3D small subvolumes extracted from tomograms. We demonstrate that the autoencoder can be used for efficient and coarse characterization of features of macromolecular complexes and surfaces, such as membranes. In addition, the autoencoder can be used to detect non-cellular features related to sample preparation and data collection, such as carbon edges from the grid and tomogram boundaries. The autoencoder is also able to detect patterns that may indicate spatial interactions between cellular components. Furthermore, we demonstrate that our autoencoder can be used for weakly supervised semantic segmentation of cellular components, requiring a very small amount of manual annotation.

## 1 Introduction

Recent developments in cellular electron cryo-tomography (CECT) have enabled the three-dimensional visualization of cellular organization in the near-native state and at submolecular resolution. Subcellular components can be systematically analyzed at unprecedented levels of detail. This *in situ* 3D visualization has made possible the discovery of numerous important structural features in both prokaryotic and eukaryotic cells as well as in viruses [17, 7, 14, 18]. As the approach develops, high quality CECT data will continue to yield valuable insights into the structural organization of the cell. In principle, a tomogram of a cell contains structural information of all cellular components within the field of view. However, cellular structures are densely packed within a small volume, which makes it challenging to systemically extract cellular structural information from tomograms. Imaging limitations, such as low signal-to-noise ratio (SNR) and missing wedge effects, further complicate the systematic recovery of such information. Currently, many CECT structural identification, characterization and segmentation tasks are performed by visual inspection and manual annotation, which can be very laborious. Consequently, the labor-intensive nature of these analyses has become a major bottleneck in CECT studies.

*Corresponding author email: mxu1@cs.cmu.edu

Structural separation approaches for macromolecules may be used for facilitating the systematic and automatic characterization of structural or image features. Inside a tomogram, a macromolecule can be extracted and represented as a *subtomogram*, which is a 3D small subvolume (3D analog of a 2D image patch) of cubic shape. Reference-free subtomogram classification [e.g. 5, 42, 11, 38] has been developed for the structural separation of macromolecules. Such approaches are designed for recovering structures of large macromolecular complexes. Nevertheless, the steps for subtomogram alignment or integration over all rigid transformations in those approaches are computationally intensive, and therefore limit the scalability of these approaches. To increase scalability, we and others have developed 3D rotational invariant feature [43, 44, 10] approaches. Recently, we developed a supervised deep structural feature extraction approach [45] that can be used for the characterization of structural or image features. Nevertheless, this method employs a supervised approach that requires data annotation for training.

In this paper, we complement existing approaches by developing an unsupervised approach for automatic characterization of tomogram features. Automatic characterization of image features (represented as 3D small subvolumes) is very useful for separating heterogeneous small subvolumes into homogeneous small subvolume sets, which simplifies the structural mining process by separating structures with different shapes or orientations. Although resulting small subvolume sets are not labeled, image feature clues are provided to guide the identification of representative structures. Unknown structures of the same type and orientation are likely to be clustered in the same small subvolume set, which helps the identification of the structure and spatial organization in a systematic fashion.

Specifically, we propose a 3D convolutional autoencoder model for efficient unsupervised encoding of image features (Figure 1a). A convolutional autoencoder is a type of Convolutional Neural Network (CNN) designed for unsupervised deep learning. The convolutional layers are used for automatic extraction of an image feature hierarchy. The training of the autoencoder encodes image features (represented as 3D small subvolumes) into compressed representations. The encoded image features are then clustered using k-means clustering. a small subvolume set is characterized by the decoded cluster center. With an optional fast preprocessing step of pose normalization, and with GPU acceleration, the separation process is significantly more scalable than the subtomogram classification [e.g. 42] and pattern mining [46] approaches. As a result, it is particularly suitable for unsupervised structural mining among large amounts of small subvolumes and identifying representative structures with representative orientations. Through testing our approach on experimental cellular cryo-tomograms (Section 3.3), we are able to efficiently encode and cluster tens of thousands of small subvolumes using a single GPU. We identified 1) surface features such as membranes, carbon edges, and tomogram boundaries of certain orientations and 2) large globular features corresponding to macromolecular complexes likely to be ribosomes. Both the surface features and the large globular features were qualitatively validated by embedding these patterns back into the tomograms. Interestingly, we further identified a spatial interaction pattern between cellular components which is difficult to identify through visual inspection of the tomogram. Moreover, we performed a numerical study on simulated data to analyze the accuracy of our autoencoder model on detecting surface features and ribosome structures, and to assess the missing wedge effect (Supplementary Section S3).

To reduce the dependence of our convolutional autoencoder model to the variation in orientation and translation of image features, as an optional step, we further adapted a pose normalization approach [46] to normalize the location and orientation of image features. After pose normalization, the image features of similar structure have similar orientation and location. Therefore, unknown structures of similar shape are more likely to be clustered in the same small subvolume set, which assists the characterization of the image features in a less orientation dependent fashion. Our tests on both experimental and simulated tomograms demonstrate the efficacy of the combination of pose normalization and convolutional autoencoder (Figure 7 and Supplementary Sections S3).

Manual segmentation of tomograms is a laborious task. To facilitate structural segmentation, automatic or semi-automatic approaches have been developed for segmenting specific structures. Such approaches use manually designed rules for 1) extraction of image features characterizing the specific ultrastructure and 2) segmentation based on combinations of extracted image features. Often, feature extraction and segmentation rules are specifically designed for particular types of image features or ultrastructures, such as membranes [e.g. 6, 28, 29, 13] or actin filaments [e.g. 37, 47]. Only very few generic and unified approaches exist for segmenting various structures [e.g. 9, 27]. Generic and unified approaches come with the advantage of being easily extended to segmenting new types of structures through automatic learning rules, which is often done in a supervised fashion. In recent years, deep learning-based approaches [e.g. 26] have emerged as dominant approaches for supervised generic segmentation in computer vision applications due to their superior performance in the presence of large amounts of training data. Deep learning has also been used for generic segmentation of ultrastructures [9] in cellular tomograms. Since supervised segmentation approaches often rely on training data prepared through manual segmentation of images, it is beneficial to develop approaches to reduce the amount of supervision (in terms of manual annotation) to facilitate the automation of training data preparation.

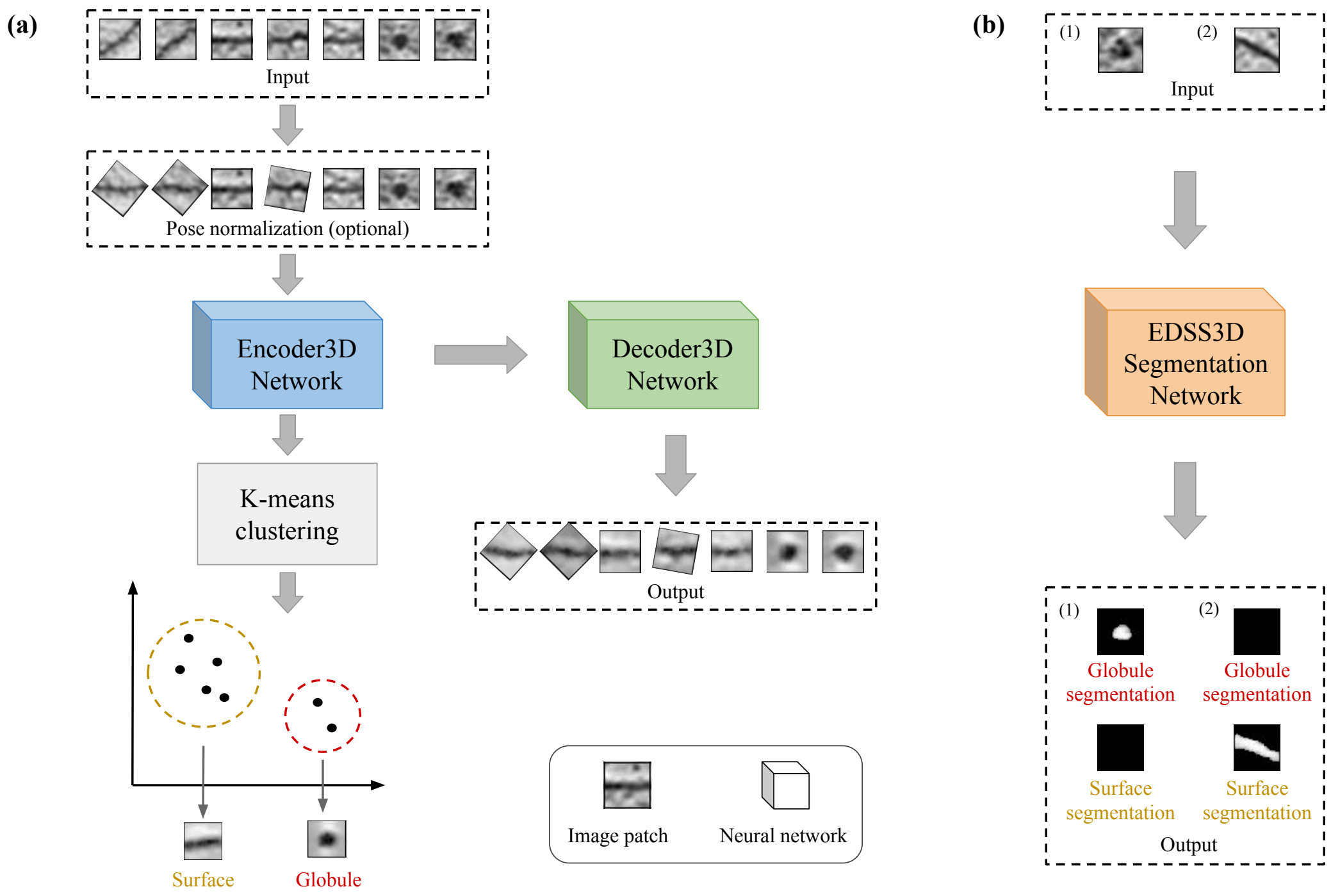

Figure 1: Conceptual diagrams of (a) Autoencoder for characterization of small subvolumes (Section 2.2). (b) Encoder-decoder network for small subvolume semantic segmentation (Section 2.4).

To complement existing approaches through reducing the amount of supervision, in this paper, we demonstrate that the cluster groups generated from our autoencoder can be used to train a 3D CNN model for semantic segmentation in a weakly supervised fashion. In particular, after simple manual selection and grouping the clusters, the cluster groups are used to train dense classifiers for the voxel-level classification and to semantically segment tomograms. In the whole segmentation pipeline, the amount of manual voxel-wise segmentation of 3D images can be dramatically reduced. The only step that requires manual intervention is the selection and grouping of image feature clusters among a number (such as 100) of candidate clusters, based on the decoded cluster centers. Therefore, the whole pipeline is weakly supervised and requires only a small amount human intervention. Our preliminary tests and qualitative assessments on experimental tomograms demonstrate the efficacy of our approach (Section 3.3).

**Our contributions are summarized as follows**:

1. We designed a deep autoencoder network for unsupervised clustering of CECT small subvolumes to provide a fast and coarse mining and selection of CECT small subvolumes without any annotated training data. Specifically, we adapted 2D autoencoder networks to 3D networks for CECT data. Also, we combined k-means clustering algorithms with autoencoder networks to provide clustering of CECT small subvolumes into sets with homogeneous image features.

2. To merge small subvolumes of similar image features but different orientation together, we adopted a pose normalization approach for normalizing the location and orientation of structures in a small subvolume. As a result, small subvolumes of similar image features with different orientations are more likely to be grouped into the same image feature cluster.

3. We designed an encoder-decoder semantic segmentation network for weakly supervised coarse segmentation of tomograms. This approach can effectively reduce the amount of manual voxel-wise segmentation of simple image features.

# 2 Methods

## 2.1 Background

Deep learning is one of the most popular computer vision techniques used today across a broad spectrum of applications [22]. Convolutional Neural Networks (CNN) [23], an artificial neural network inspired by the hierarchical organization of animal visual cortex, have achieved high performance and accuracy in computer vision tasks such as image classification [e.g. 21] and semantic segmentation [e.g. [26]]. A CNN model is a combination of layers in sequence and each layer consists of a certain number of neurons with receptive fields on the previous layer. In this paper, we employ CNN to encode CECT small subvolumes to low dimensional vectors for clustering. The use of a stack of convolution layers has the advantage of learning the inherent structure of local correlations and hierarchical organization of correlations in images. The details of different types of CNN layers, activation functions, and the optimization techniques are introduced in Supplementary Section S1.

## 2.2 Autoencoder3D network for unsupervised image feature characterization

A typical autoencoder [16] consists of two main components, the encoder $\phi$: $X \to F$, which encodes the input $X$ to a representation $F$, usually in the form of low dimensional vector, and the decoder $\psi$: $F \to \hat{X}$, which decodes the $F$ to a reconstruction of $X$, $\hat{X}$. The autoencoder network is trained to minimize the difference between input $X$ and reconstruction output $\hat{X}$. Normally, the goal of the autoencoder network is to reduce the dimension of input and to characterize image features with high precision.

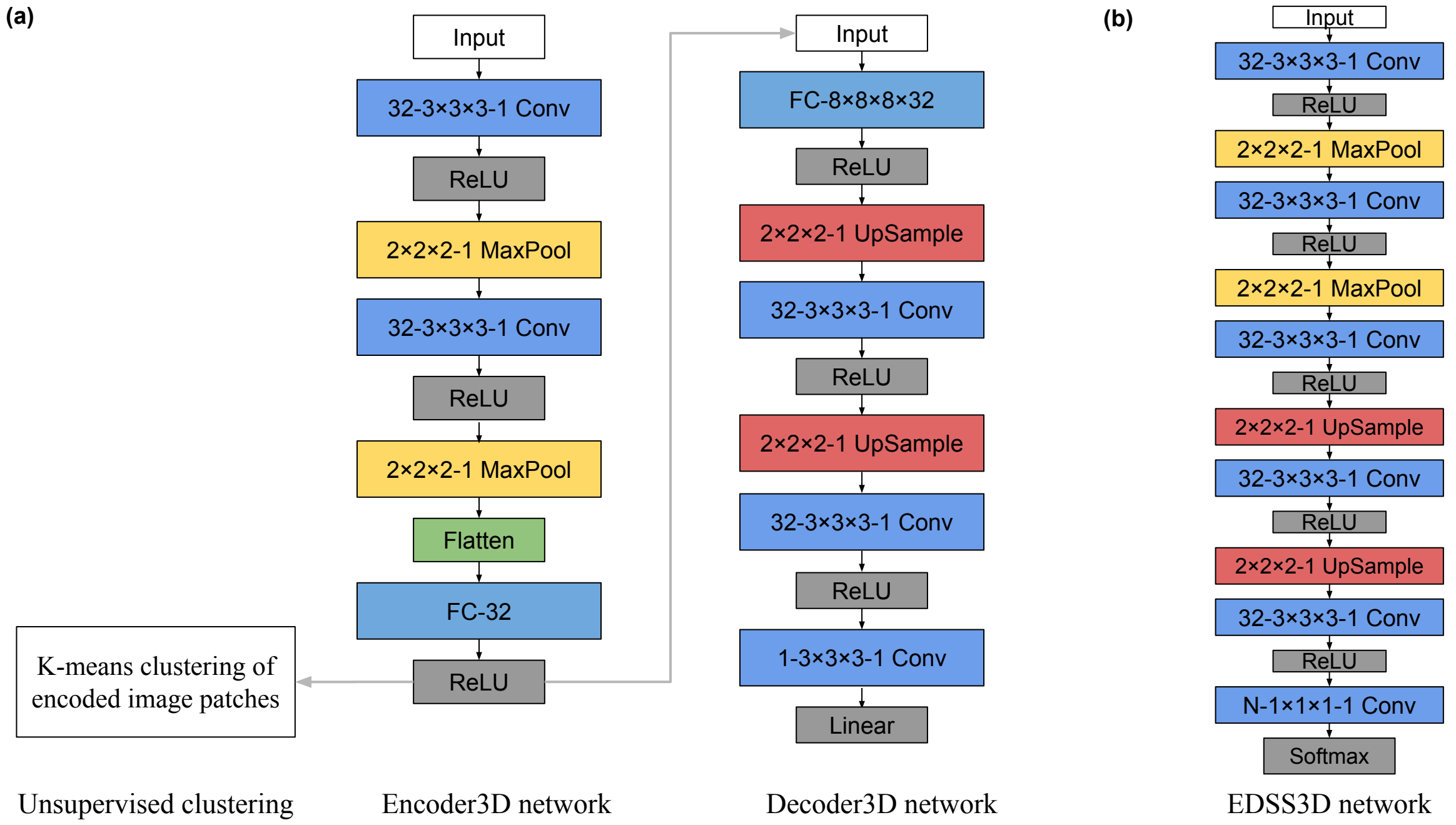

Figure 2: The network architecture of our CNN models. All three networks have multiple layers. Each colored box represents one layer. The type and configuration of each layer are shown inside each box. For example, '$32-3\times3\times3-1$ Conv' denotes a 3D convolutional layer with 32 filters, $3\times3\times3$ kernel size, and 1 stride. '$2\times2\times2-1$ MaxPool or UpSample' denotes a 3D max pooling or upsampling layer over a $2\times2\times2$ region with 1 stride, respectively. 'FC-$8\times8\times8\times32$' denotes a fully connected layer with neurons of size $8\times8\times8\times32$, where every neuron is connected to every output of the previous layer. 'Flatten' denotes a layer that flattens the input. $N$ is the number of classes in the semantic segmentation training set. 'ReLU', 'Linear', 'Softmax' denote different types of activation layers. See Supplementary Section S1 for details of different layers.

We propose a 3D convolutional autoencoder model, denoted as Autoencoder3D. Our model consists of four types of layers: convolution, pooling, fully connected, and softmax (Supplementary Section S1 for details). The body of standard CNN models for computer vision tasks is designed to have alternating convolutional layers and pooling layers. We adopted such design into our Encoder3D network. Following standard convolutional autoencoder models, we use fully connected layers to encode the features extracted from previous layer into a 32-dimensional vector. Since the Encoder3D network encodes an input small subvolume to a 32-dimensional vector and the Decoder3D network decodes the encoded vector to reconstruct the input image, the architecture of Decoder3D network is a mirror reversal of the Encoder3D network, with up-sampling layers replacing max-pooling layers. The input of Autoencoder3D network is a 3D small subvolume extracted from a tomogram, represented as a 3D array $A$ of $\mathbb{R}^{m \times n \times p}$. The Encoder3D network encodes the small subvolume $A$ as an encoding vector $v$ of $\mathbb{R}^{32}$. The Decoder3D network decodes the encoding vector $v$ to a reconstruction $\hat{A}$ of the same size $\mathbb{R}^{m \times n \times p}$.

The architecture of the Autoencoder3D model is shown in Figure 2a. The Encoder3D part contains two convolutional layers with $3 \times 3 \times 3$ 3D filters, two $2 \times 2 \times 2$ 3D max pooling layers, and one fully connected output layer outputting vector $v$ of length 32. We use $L1$ norm regularization to encourage sparsity in the encoded features. Previous work [31] shows that sparsity regularization improves autoencoder performance. The Decoder3D part contains one fully connected layer with the same output shape as the input shape of the Encoder3D fully connected output layer, two convolutional layers with $3 \times 3 \times 3$ 3D filters, two $2 \times 2 \times 2$ 3D upsampling layers, and one convolutional output layer with $3 \times 3 \times 3$ 3D filters. All hidden layers and the Encoder3D fully connected output layer are equipped with the rectified linear (ReLU) activation. The Decoder3D convolutional output layer is equipped with a linear activation.

Simple CNNs with less number of layers and neurons are faster to train and compute. Increasing the number of layers and neurons may increase the capacity and accuracy of predictions, but can be prone to issues such as over-fitting. Over-fitting occurs when a model fits too close to the training data that cannot perform well for testing data. We have tried different autoencoder networks with a range of layer numbers and neuron numbers. We designed the current CNN networks by balancing the efficiency, validation accuracy, and training speed. The user can directly apply our networks to their CECT small subvolumes data. Our networks were optimized to have high validation accuracy and fast training speed. The networks can be further optimized based on the same guideline.

The autoencoder follows a similar principle as sparse coding [25]. Studies have shown that a small subvolume can be effectively represented by a linear combination of a small number of basis vectors [25]. When designing our Autoencoder3D model, we have tried different encoding vector length, from 16, 32, to 128. By visually comparing the decoded images patches with the corresponding input small subvolumes, we observed that, when using encoding vector of length 32, several elements in the encoding became zeros for all small subvolumes. Increasing the encoder vector length further will not change the results much. A 32-dimensional encoding vector was enough for our task. The rationale behind the deep autoencoder is that images can be compressed to a very simple vector, which can be decoded to reconstruct the original image. An element in the encoding vector does not mean to encode only one feature. Since there are millions of parameters in the decoding network, 32 elements in combination can already encode and produce a large number of features.

The Encoder3D network contains two $2 \times 2 \times 2$ 3D max pooling layers whereas the Decoder3D network contains two $2 \times 2 \times 2$ 3D upsampling layers. Therefore, to have the subvolume reconstruction output the same size as the input, all three dimensions of the input subvolume must be extracted to be a multiple of 4. For example, after two rounds of $2 \times 2 \times 2$ max pooling, a subvolume of size $40^3$ will become an array of size $10^3$. Then, a flattening layer is applied, which will flatten the $10^3$ array to a one dimensional vector of length 1000. Then a fully connected layer, which can take inputs of arbitrary length, will process the one dimensional vector of length 1000 to be an encoding of length 32. This encoding will be used for clustering. Every neuron in the fully connected layer is connected to all 1000 elements of the input vector. This is why a fully connected layer can take inputs of arbitrary length. We note here the encoding vector length 32 is not related to the size of the input subvolume.

## 2.3 Unsupervised learning for grouping of small subvolumes

Clustering is a necessary step for collecting relatively homogeneous groups of small subvolumes from heterogeneous inputs. However, due to the high dimensionality of the samples, it is extremely difficult to discriminate two small subvolumes only based on simple distance measures [45, 3]. Therefore, we propose an unsupervised small subvolume clustering approach based on encoded features of substantially lower dimensions. Using the Autoencoder3D network, each small subvolume is encoded into a vector of real numbers that represent features of the original small subvolume. K-means clustering is then applied to group similar small subvolumes together based on the encoding.

We note here that after k-means clustering, a simple step of manually selecting interested clusters is needed to further supervise semantic segmentation of new datasets in Section 2.4. So, the decoded cluster centers are plotted to guide the user to select and group clusters of interest. Selected and grouped clusters are used as positive samples in a dataset for training a semantic segmentation model defined in Section 2.4. An example of selecting and semantic segmentation model training is described in 3.3 and 3.4. The segmentation used for training in the training set is obtained by thresholding the decoded 3D images at a certain mask level.

## 2.4 EDSS3D network for weakly supervised semantic segmentation

In this section, we propose a 3D encoder-decoder semantic segmentation network (EDSS3D) to perform supervised segmentation of new small subvolume data based on previous unsupervised learning results. The design of the model is inspired by [4]. The input of EDSS3D network is a 3D small subvolume, represented as a 3D array $B$ of size $\mathbb{R}^{m\times n\times p}$, extracted from a testing dataset tomogram. The EDSS3D network outputs $L$ number of 3D arrays $B'_l$ of the same size $\mathbb{R}^{m\times n\times p}$, where $L$ is the number of semantic classes and each voxel in $B'_l$ denotes the segmentation probability of this voxel belonging to the $l^{\text{th}}$ semantic class.

In particular, the decoded 3D images of selected clusters are used as training data. The architecture of EDSS3D model is shown in Figure 2b. The architecture consists of five convolutional layers with $3\times 3\times 3$ 3D filters, two $2\times 2\times 2$ 3D max pooling layers, two $2\times 2\times 2$ 3D upsampling layers, and one convolutional 3D output layer with the number of filters equal to the number of segmentation classes. All hidden layers are equipped with ReLU activation layer. The convolutional 3D output layer is equipped with a softmax activation layer.

We adopted the standard CNN model design into our EDSS3D network. Similar to the Autoencoder3D model, our EDSS3D model is an encoder-decoder bottleneck-type model with same-size output as the input for each class. However, the Autoencoder3D model performs image encoding for unsupervised image feature characterization whereas EDSS3D model performs supervised image semantic segmentation. Accordingly, the EDSS3D model is not broken into two parts. Also, for multi-class classification, the EDSS3D has an output softmax activation layer rather than a linear activation layer.

## 2.5 Optional preprocessing step: pose normalization of small subvolumes

CECT small subvolumes contain image features of different orientations. The similar image feature of different orientations often cannot be clustered together. Previously, we have developed level set based pose normalization for pre-filtering of subtomograms [46]. We adapted this method as an optional step for preprocessing small subvolumes by directly normalizing the orientation and location of image features.

Specifically, before the small subvolumes are used to train the Autoencoder3D model, the center of mass and principal direction of each small subvolume are calculated. The principal direction of a small subvolume is computed as the directions of the first two principal components in principal component analysis [41]. Each small subvolume is translated and rotated according to its center of mass and principal directions. Therefore, the orientation and location of a feature inside a small subvolume are normalized for better clustering. Some voxel values of a rotated and translated small subvolume may be missing due to rotation and translation operation. Those missing values are filled using the corresponding image intensities from the original tomogram.

In a small subvolume, voxels with negative values of high magnitude correspond to regions with high electron density. Before pose normalization, we normalize small subvolume values so that all values are positive and the signal regions have higher values. We note here that the value normalization is only used for calculating the center of mass and the principle component. Rotated and translated small subvolumes will still have negative values of high magnitude corresponding to regions with high electron density. Let $\vec{r} = [x,y,z]^\top$ denote the locations of voxels in a small subvolume and $f(\vec{r})$ denote the normalized value at location $\vec{r}$. First, we calculate a center of mass $\vec{c}$ of $x$:

$$\vec{c} = \frac{\int_{\vec{r}} f(\vec{r})\,\vec{r}}{\int_{\vec{r}} f(\vec{r})} \tag{1}$$

.

Then, we calculate

$$W = \int_{\vec{r}} [f(\vec{r})]^2 (\vec{r}-\vec{c})(\vec{r}-\vec{c})^\top \tag{2}$$

.

We apply the eigen decomposition $W = Q\Lambda Q^T$ of $W$ where $Q$ is an orthogonal matrix of eigenvectors and $\Lambda$ is a diagonal matrix of eigenvalues in descending order by their magnitude. Pose normalization is performed on a small subvolume by first translating the center of mass to the center and then rotating the small subvolume using $Q$ as the rotation matrix. Examples of pose normalization on surface small subvolumes are shown in Figure 3. After pose normalization, the surface small subvolumes are normalized to be horizontal orientation located in the center.

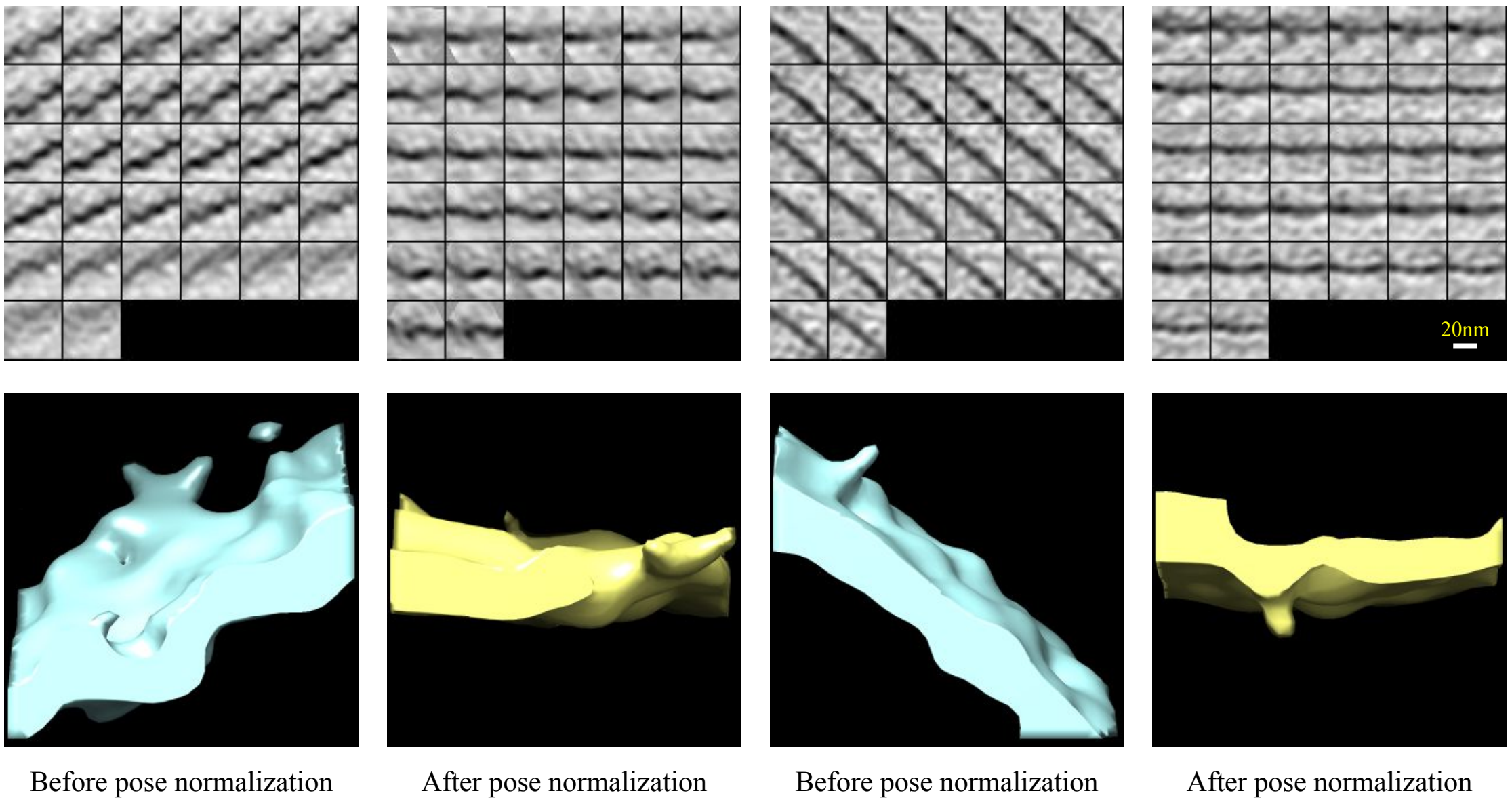

Figure 3: Examples of surface small subvolumes (2D slices) before and after pose normalization. For better visualization, small subvolumes are Gaussian smoothed with $\delta = 2.0$. See Supplementary Section S2 for details of Gaussian smoothing. Isosurface views are plotted below the 2D slices.

Remark: Convolutional neural networks can input image of arbitrary size. Therefore, it is a potential advantage of handling non-cubic small subvolumes. However, if the optional pose normalization preprocessing step is applied, it is preferable for the input small subvolumes to be cubic shape to facilitate rotation operation.

### 2.6 Implementation details

The training and testing of our CNN models were implemented using Keras [12] and Tensorflow [2]. Image processing suite EMAN2 was used for reading tomograms [39]. A variant of our Tomominer library was used for data preparation and image display [15]. K-means clustering was performed using the Sklearn toolbox [33]. Chimera [35] and Mayavi [36] were used to generate the embedded tomogram figures. The experiments were performed on a computer equipped with Nvidia GTX 1080 GPU, one Intel Core i5-5300U CPU, and 128 GB memory.

## 3 Results

### 3.1 Acquisition of experimental tomograms

To test our implementation, we used two cellular tomograms of COS-7 (*Cercopithecus aethiops* kidney) cells. Cells were grown on c-flat gold mesh carbon-coated holey carbon grids to a density of 1-2 cells/grid square. Cells were maintained at 37°C, 5% $CO_2$ in Dulbecco's modified Eagle's medium supplemented with L-glutamine, nonessential amino acids, and 10% fetal bovine serum. Prior to freezing, BSA-conjugated 10-nm gold fiducial markers were added to grids, which were then blotted manually from the backside for 4 s, and plunged into a liquid ethane/propane mixture cooled to liquid $N_2$ temperature. Tilt series were collected on a Tecnai TF30 "Polara" electron microscope equipped with a Quantum postcolumn energy filter (Gatan) operated in zero-loss imaging mode with a 20-eV energy-selecting slit. All images were recorded on a postfilter $\approx 4000 \times 4000$ K2-summit direct electron detector (Gatan) operated in counting mode with dose fractionation. Tilt series were collected using SerialEM at a defocus of -6 $\mu$m. Tilt series covered an angular range of $-60^{\circ}$ to $+50^{\circ}$ in increments of $4^{\circ}$. Tomograms were reconstructed in IMOD using weighted back-projection, with a voxel size of 0.355 nm. The tomograms are not collected for the purpose of subtomogram averaging, therefore they are not CTF corrected. They were further binned four times to reduce size. The resulting two tomograms were termed COS-7 tomogram 1 and COS-7 tomogram 2.

### 3.2 Data preparation and autoencoder training

To collect small subvolumes, we performed a template-free Difference of Gaussian (DoG) particle picking process as described in [34]. The COS-7 tomogram 1 was convolved with a Gaussian Kernel of $\sigma = 2$ voxelmissins in radius for smoothing and then with a 3D DoG function with scaling factor of $\sigma = 5$ voxels in radius and scaling factor ratio $K = 1.1$ for small subvolume extraction. Potential macromolecules detected as peaks in the DoG map were filtered so that the distance between peaks were at least 10 voxels. 38112 small subvolumes of size $32^3$ voxels were extracted for autoencoder network training. In principle, one can also use a sliding window to extract small subvolumes. However, a sliding window on a 3D image would produce a substantially larger amount of small subvolumes that would introduce a substantially larger amount of computational burden.

We randomly split the 38112 small subvolumes into a training set of size 34300 and a validation set of size 3812. The Autoencoder3D model was trained using optimizer Adaptive Moment Estimation (Adam) with exponential decay rates $\beta_1 = 0.9$ and $\beta_2 = 0.99$ to minimize the mean squared error loss function [20] (see Supplementary Section S1 for details of Adam). After one epoch training, the model was saved only if there was an improvement in validation dataset loss compared to the previous epoch. Adam training was performed with learning rate 0.001 and a batch size of 8 until the validation dataset loss did not improve for 20 consecutive epochs. The learning rate in CNNs defines the "step size" of a gradient update. When the learning rate is too high, such as 1, the output truth over-corrects the model output and overshoots the optima that we are trying to converge to, which will make the CNN training highly unstable. When the learning rate is too low, steps are too small and training will take much longer to converge. To find the optimal learning rate, we started with an initial value of 0.1, and decreased it until the training began to stably converge.

We measured the computation speed of Autoencoder3D network. On average, the training took 0.013s per small subvolume per epoch for Autoencoder3D. Given the trained Autoencoder3D model, on average, the encoding of a small subvolume took 0.0012s and the decoding of an encoded small subvolume took 0.0023s. Therefore, our model can be used to quickly process large amounts of small subvolumes. Computing times and environments of all steps can be found in Supplementary Table S2.

### 3.3 K-means clustering of encoded features

After training, the Autoencoder3D network was used to encode each small subvolume as a 32-dimensional vector. Then, we performed k-means clustering with $k = 100$ on the encoded small subvolumes to group similar small subvolumes together. The cluster center of each group, a 32-dimensional vector, was decoded to a 3D small subvolume reconstruction by the Decoder3D network. Figure 4 shows examples of decoded cluster centers.

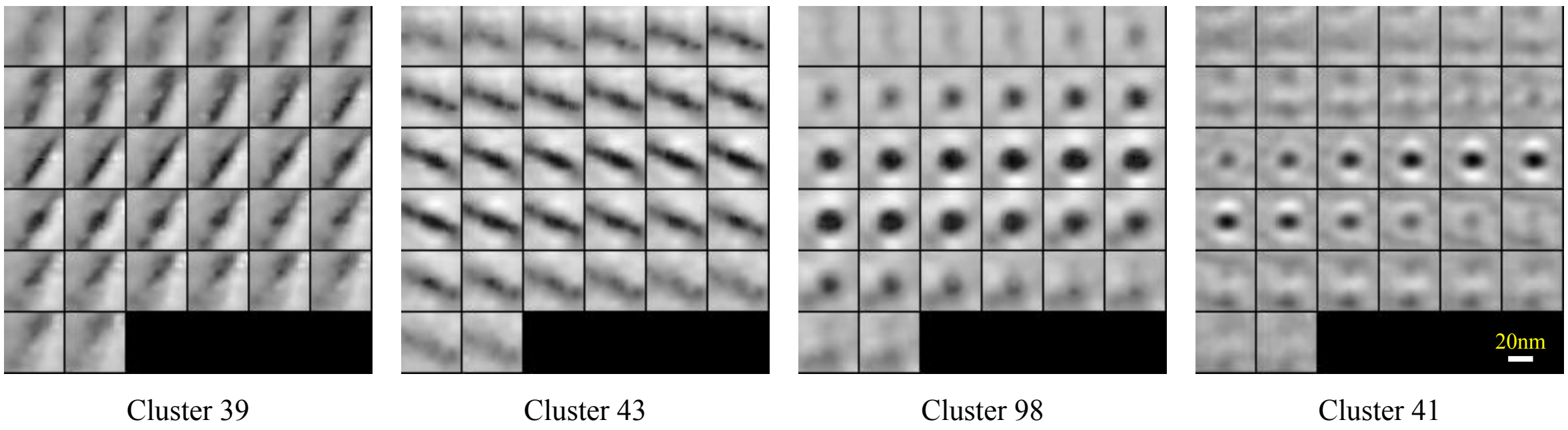

Figure 4: Four decoded cluster centers obtained from COS-7 tomogram 1. For each cluster center, 2D slices of the decoded 3D images are shown. The slices are 32 images representing $32^2$ voxels, representing a decoded 3D image of size $32^3$. Cluster 39 and 43 are selected surface feature clusters. Cluster 98 is a selected large globular feature cluster. Cluster 41 is an example of a non-selected cluster that contains small globular image feature. The majority of non-selected clusters look like this small globular image feature.

It is evident that clusters 39 and 43 represent parts of surface fragments seen in different orientations. Cluster 98 represents globular macromolecules with sizes similar to that of established ribosomal macromolecules; as such the characteristic structures contained in this cluster are likely ribosomes (termed ribosome-like structures). A further inspection of 'large globule' small subvolumes by template searching and reference-free subtomogram averaging can be found in Supplementary Section S4. The majority of non-selected classes look like 'small globule' structure as in cluster 41.

After manually labeling these 100 clusters, we selected 10 clusters of 500 small subvolumes that represented surface features of different orientations and 7 clusters of 308 small subvolumes that represented large globular features. The total 808 small subvolumes of surface and large globular features were used to annotate the COS-7 tomogram 1 (Figure 5). In Figure 5, parts of membranes, carbon edge, and tomogram boundary regions are automatically annotated based on our cluster results. Large globular features are annotated across a large region in the tomogram.

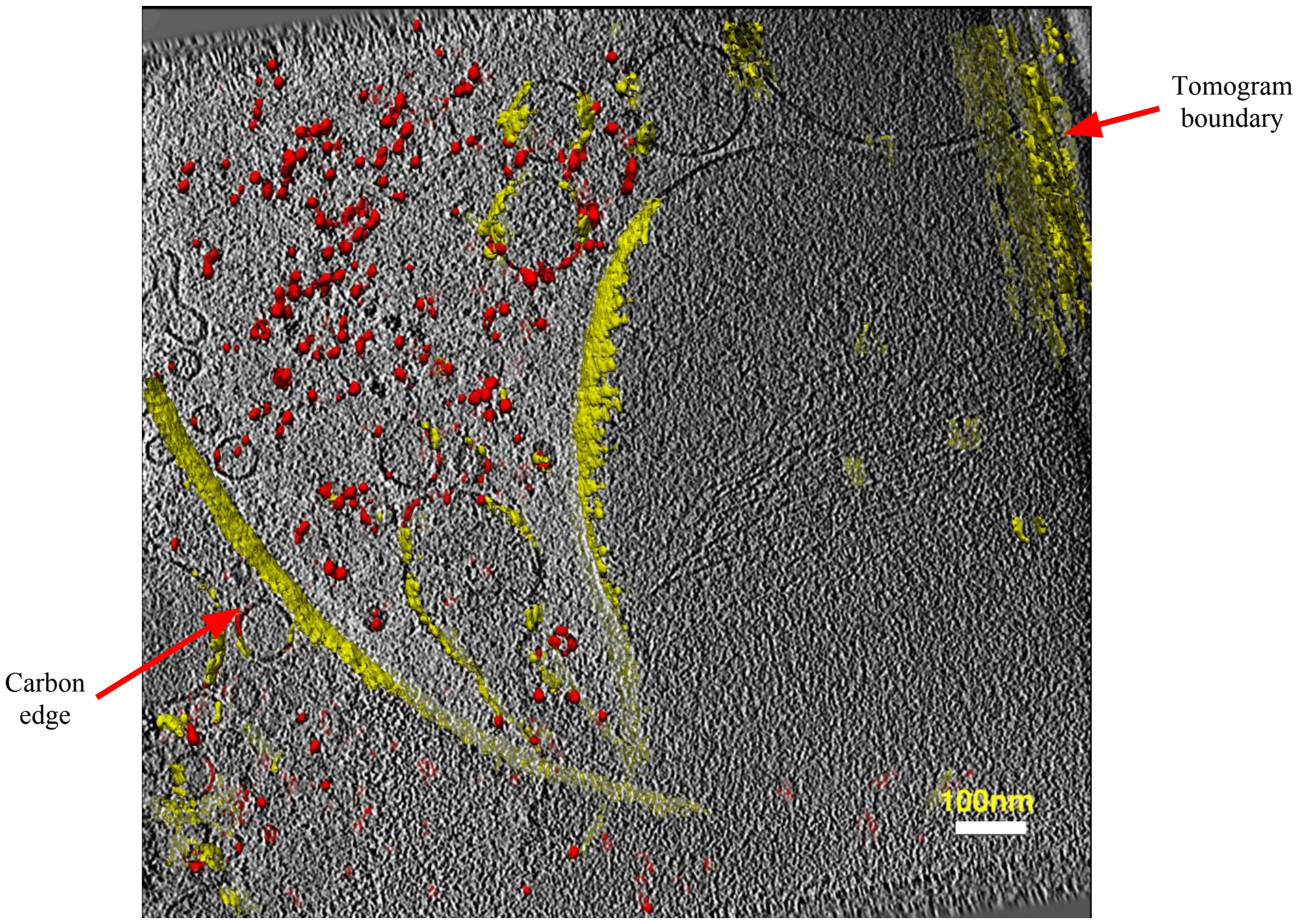

Figure 5: Isosurfaces of decoded small subvolumes of selected clusters embedded to the COS-7 tomogram 1. Surface features (yellow) and large globular features (red) are annotated in the tomogram. A long carbon edge and a tomogram boundary, annotated in yellow, are indicated by red arrows.

We also explored the impact of pose normalization on model training and clustering. After the small subvolumes were extracted from the COS-7 tomogram 1, images patches values were normalized by taking the inverse and subtracting the minimum value. Each small subvolume was then pose normalized (Section 2.5) and used for Autoencoder3D training (Section 3.2). Then, the encoded small subvolumes were clustered using k-means clustering with $k = 100$. Figure 6 shows examples of decoded cluster centers of surface features. After pose normalization, surface features of different orientations were normalized to the horizontal orientation, which made it easier to cluster surface features even if of different orientations.

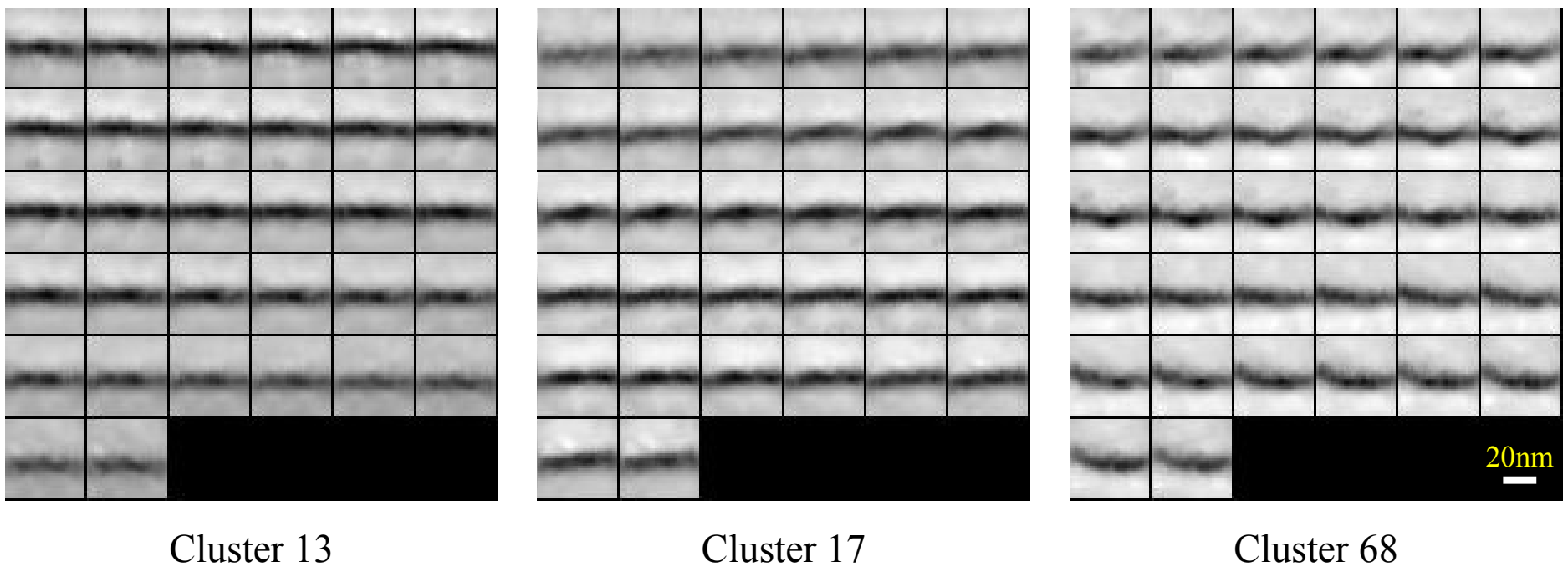

Figure 6: Three decoded cluster centers of surface features. Surface features of different orientations were pose normalized to the horizontal orientation. All 12 cluster centers of surface features resulted are of the horizontal orientation as these three clusters.

We selected 12 clusters of 900 small subvolumes that represented surface features of different orientations and 5 clusters of 370 small subvolumes that represented large globular features. The total 1270 small subvolumes of surface and large globular features were used to annotate the COS-7 tomogram 1 (Figure 7). We note here that after pose normalization, more surface features and large globular structures are selected and annotated. This could be due to the fact that pose normalization improves the clustering ability to group features of different orientations. In Figure 7, differently oriented surface features are clearly annotated.

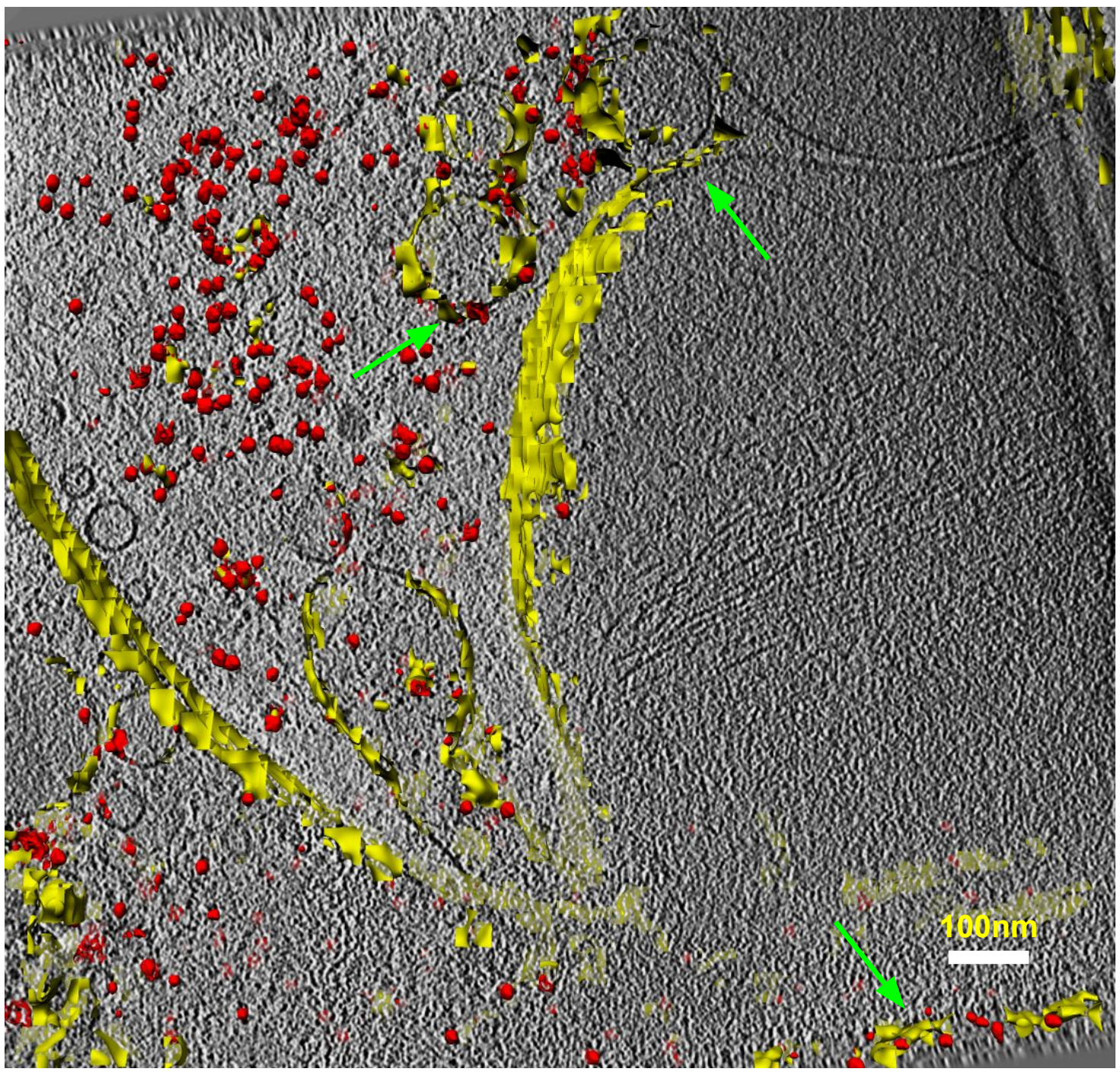

Figure 7: Isosurfaces of decoded pose normalized small subvolumes of selected clusters embedded to the COS-7 tomogram 1. Surface features (yellow) and large globular features (red) are annotated in the tomogram. Green arrows indicate horizontal surface regions that were not detected without pose normalization.

**Image features that may indicate spatial interaction** Interestingly, we detected small subvolumes that may indicate spatial interactions between cell components. By visual inspection of the location of the spatial interaction patterns, we found clusters 6, 64, and 85 to represent macromolecules that are enriched in membrane-proximal regions (Figure 8). We averaged the original small subvolumes of each of the three clusters. Figure 8 shows the 2D slices of the averaged small subvolume of the three clusters of such spatial interaction pattern (Figure 8). We are able to identify a macromolecule in the middle and some spatial interaction (likely to be membrane and macromolecule associations). To better visualize the averaged small subvolumes of this spatial interaction pattern between membrane and macromolecule, the 2D slices of images with Gaussian smoothing of $\sigma = 3$ are shown in 8. These clusters present clear evidence of such spatial interaction. The decoded cluster centers of these three clusters are also plotted, which provide additional evidence of spatial interaction enriched in membrane-proximal regions. The Gaussian smoothed averages are very consistent with the decoded cluster centers. Such consistency provides a strong evidence of the fidelity of the decoded cluster centers in representing the small subvolumes of the corresponding clusters. The validity and biological implication of this pattern remain to be further investigated.

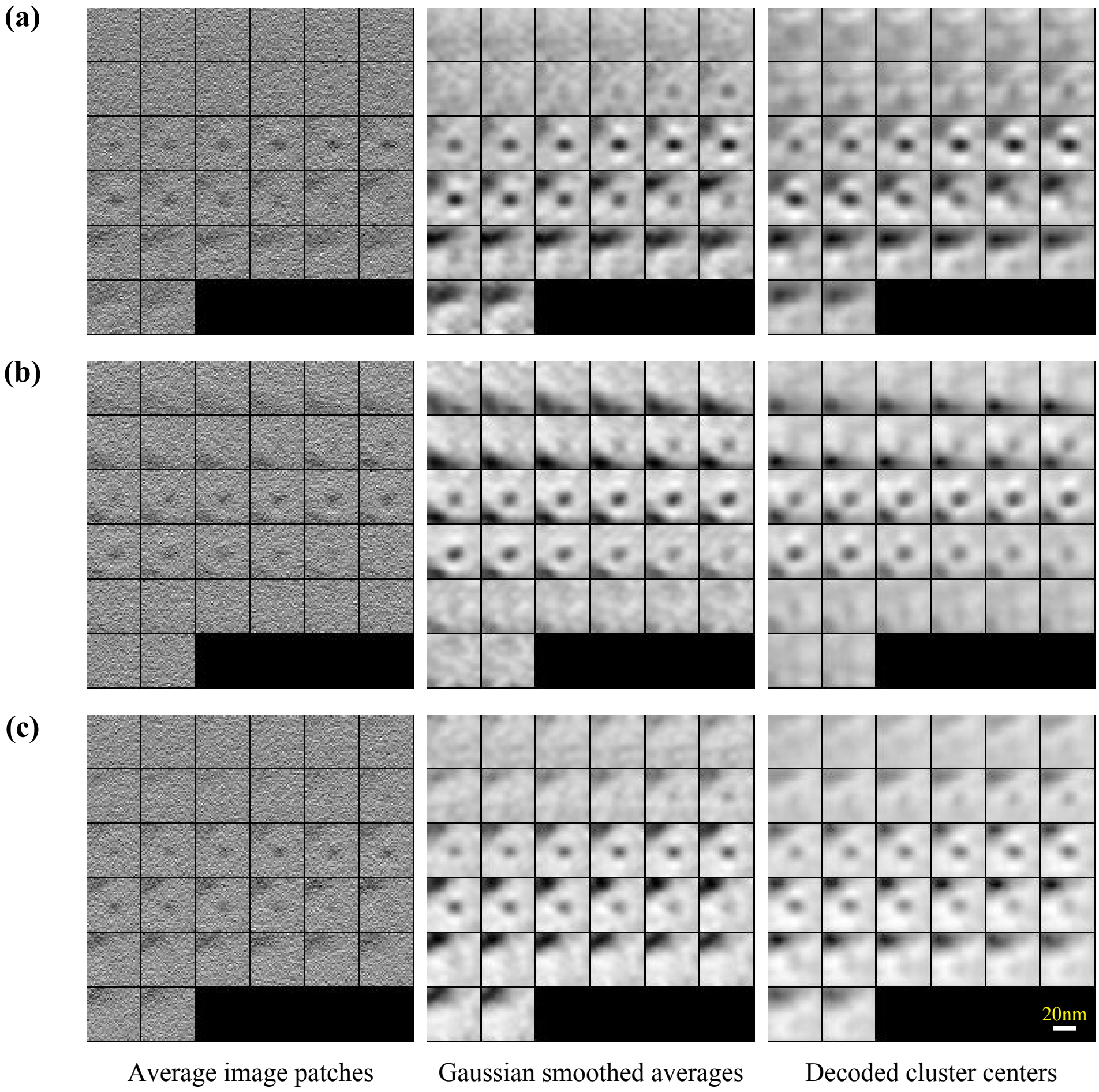

Figure 8: The 2D slices of average small subvolumes of spatial interaction pattern detected in COS-7 tomogram 1. Gaussian smoothed averages of $\sigma = 3$ are shown in the middle. The decoded cluster centers are shown on the right.

### 3.4 Semantic segmentation

**Construction of testing dataset from COS-7 tomogram 2:** A similar data preparation procedure was carried out on COS-7 tomogram 2. 42097 small subvolumes of size $32^3$ voxels were extracted. The small subvolumes were then filtered to reduce the probability of obtaining false-positive results. The 42097 small subvolumes were encoded by the trained Encoder3D network from Section 3.2. The 42097 encoded small subvolumes were mapped to its nearest cluster centroid from Section 3.3. Only the 312 small subvolumes mapped to surface feature clusters or large globular feature clusters were kept for semantic segmentation. Encodings mapped to other clusters were filtered out as they were less likely to contain any surface or large globular feature. Such filtering may also be performed through our recently developed 3D sub-volume classification approach [45].

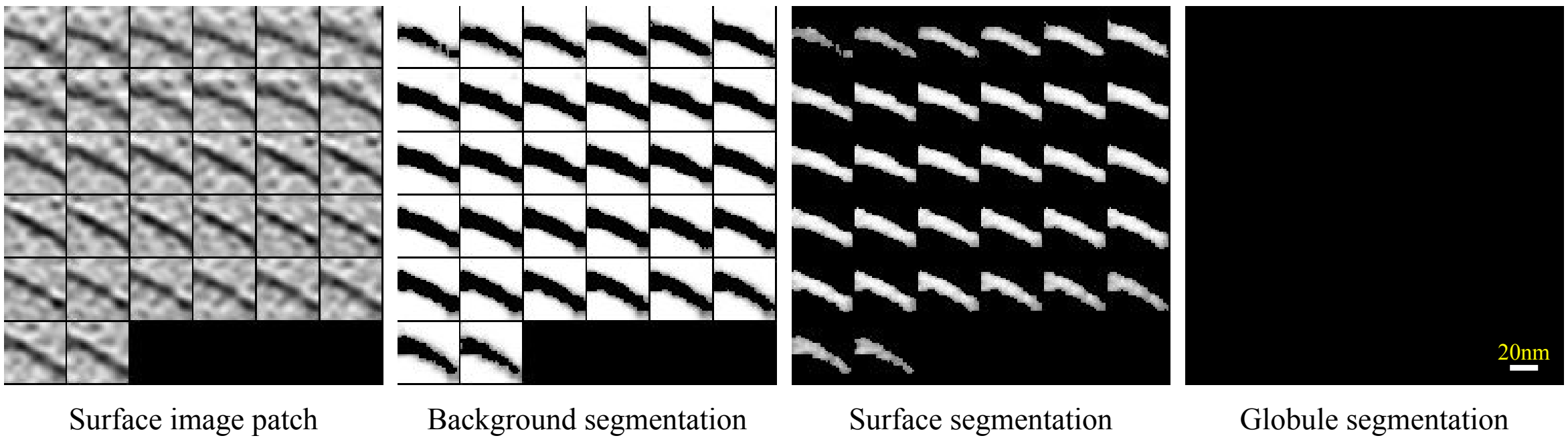

Figure 9: 2D slices of an example small subvolume (in COS-7 tomogram 2) being segmented to a surface fragment.

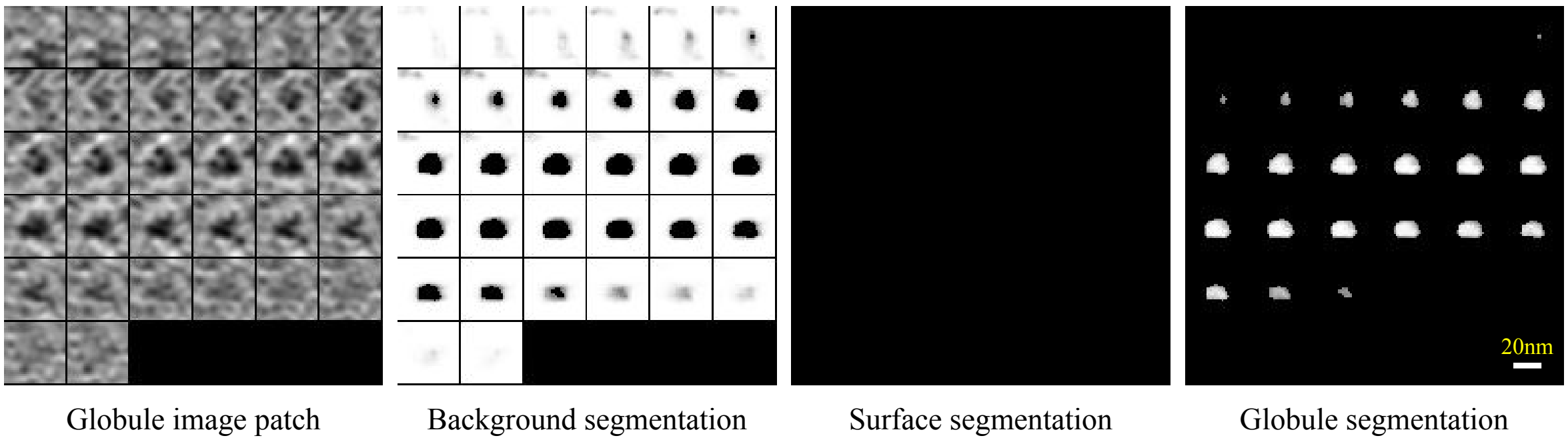

Figure 10: 2D slices of an example small subvolume (in the COS-7 tomogram 2) being segmented to a large globule.

**Construction of training dataset from COS-7 tomogram 1:** We used the k-means clustering results as a training dataset for the Encoder-decoder semantic segmentation network (EDSS3D) and then applied the trained EDSS3D network on the testing dataset.

First, the 100 decoded cluster centers were manually labeled with the two most recognizable cellular structures: surface features (membrane, carbon edge, or tomogram boundary) and electron-dense structures with the same general appearance as large globular features (termed ribosome-like structures). These structures were grouped into two classes for training. The surface feature class consisted of 10 clusters with 500 small subvolumes in total. And the large globular feature class consisted of 7 clusters with 308 small subvolumes in total. We added a third class, the background class, to denote the background regions where there was no target structure present. The segmentation ground truth was obtained by masking each decoded small subvolume in the training dataset with image intensity level 0.5. Voxels with signal greater than 0.5 were segmented as the background region and voxels with signal less than or equal to 0.5 were segmented as either the surface region or large globular region as determined by the cluster label.

**Training:** We randomly split the 808 small subvolumes into a training set of size 727 and a validation set of size 81. The encoder-decoder network model was trained using Adam with exponential decay rates $\beta_1 = 0.9$ and $\beta_2 = 0.99$ to minimize the categorical cross-entropy loss function. After one epoch training, the model was saved only if there was an improvement in validation dataset loss compared with previous epoch. Adam training was performed with learning rate 0.001 and a batch size of 128 until the loss for validation dataset did not improve for 20 consecutive epochs.

**Segmentation:** The trained EDSS3D network was applied to the testing dataset of 312 small subvolumes. 2D slices of the original testing small subvolumes and the resulting three class segmentation probability results were plotted. Figure 9 and 10 show the segmentation of two example small subvolumes. An overall visual inspection of the segmentation results on the test dataset shows that our unsupervised Autoencoder3D network and weakly supervised EDSS3D network can successfully segment this dataset into semantically meaningful classes and structures.

Figure 11 shows an embedding of segmented small subvolumes to COS-7 tomogram 2. In figure 11, some membrane regions, including many vesicular membranes, are successfully segmented and annotated in yellow. Carbon edge and tomogram boundary regions were also segmented as surface regions in yellow. Large globular features that may indicate ribosome-like structures are segmented and annotated in red across a large region in the tomogram. We note here that only a small number of small subvolumes were selected after the filtering for semantic segmentation. Some false-negative results were obtained due to the filtering. However, of the selected small subvolumes, surface regions and large globular macromolecules were successfully segmented.

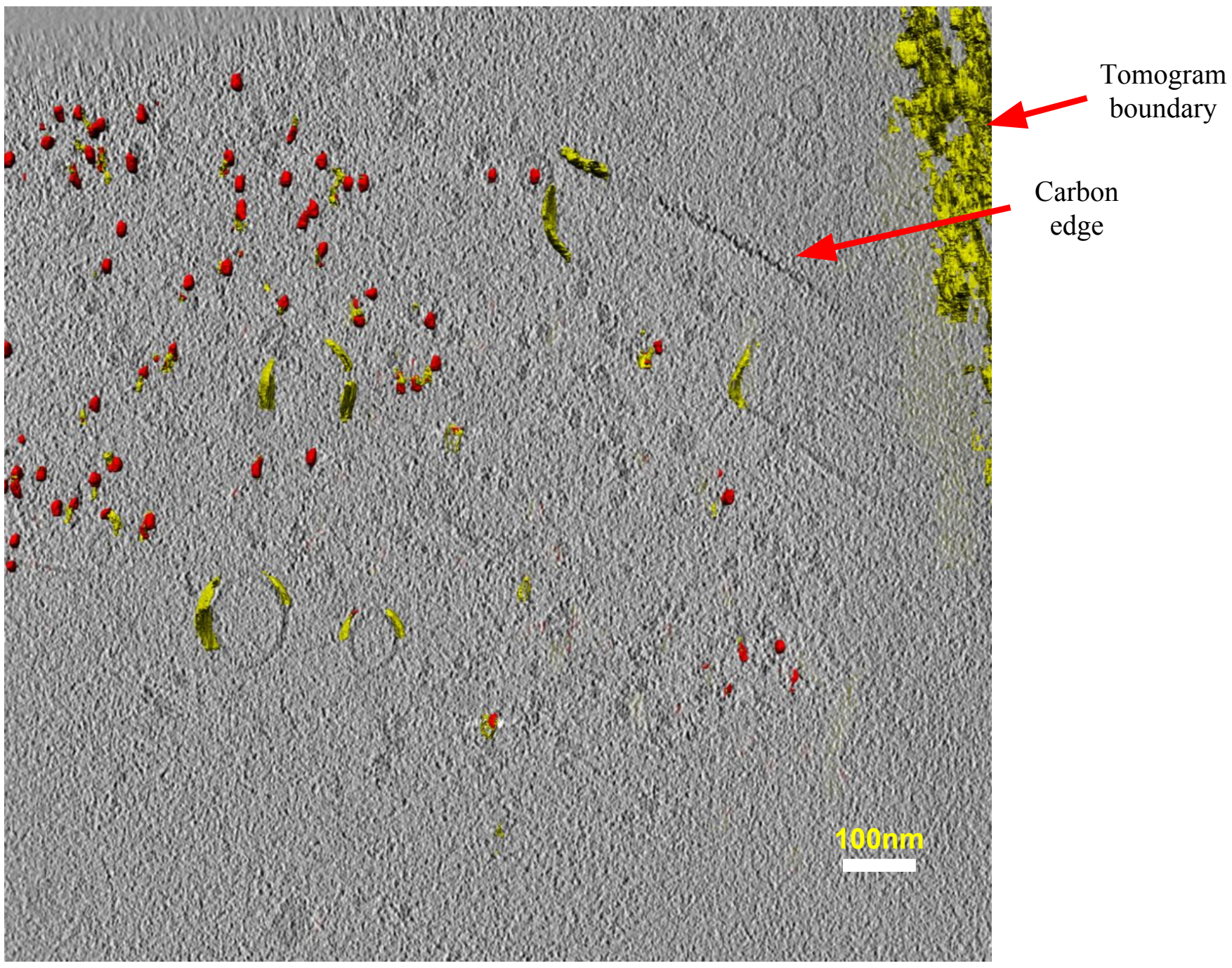

Figure 11: Annotated COS-7 tomogram 2 based on segmentation. Surface feature segmentation (yellow) and large globular macromolecules segmentation (red) are annotated in the tomogram. A tomogram boundary, annotated in yellow, and a long carbon edge, partly annotated in yellow, are indicated by red arrows.

We measured the computation speed of the EDSS3D network. On average, the training took 0.011s per small subvolume per epoch. Given the trained EDSS3D model, the segmentation took 0.011s for one small subvolume. Therefore, our model can be used to quickly segment large mount of small subvolumes. Computing times and environments of all steps can be found in Supplementary Table S2.

# 4 Discussion

CECT has emerged as a powerful tool for 3D visualization of the cellular organization at submolecular resolution and in a near-native state. However, the analysis of structures in a cellular tomogram is difficult due both to the high complexity of image content and imaging limits. To complement existing approaches, in this paper, we proposed a convolutional autoencoder approach for the fast and coarse characterization of image features among small subvolumes. We further proposed a weakly supervised semantic segmentation approach by combining the convolutional autoencoder and the full convolutional network, which only involves a very small amount of manual annotation. The preliminary tests of our approaches on experimental and simulated tomograms demonstrate the efficacy of our approaches. This proof-of-principle work presents a useful step towards the automatic and systematic structural characterization in cellular tomograms. To our knowledge, our work is the first application of convolutional autoencoder and CNN based weakly supervised semantic segmentation to the analysis of CECT data. Our methods are useful complements to current techniques.

Potential uses of our method include facilitating *in silico* structural purification and pattern mining in tomograms of intact cells, cell lysates [19], or purified complexes, and using selected image feature clusters of the same type (but possibly in different orientations) to train semantic segmentation. Moreover, template search approaches can also be facilitated by our method. In cases where the user is looking for a specific structure matched to a template, the user can ignore the resulting small subvolume sets whose cluster centers are vastly different from the template. In addition, once the feature clusters are obtained by our Autoencoder3D network, they can then be used to extract, recognize, filter, or to enhance specific types of image features. Selected small subvolume clusters grouped by different types of features can be directly used to train a classifier to recognize these features in a similar way as our recent work [45].

The main motivation of our use of convolutional autoencoder is as follows: First, direct classification (clustering) of 3D CECT small subvolumes is challenging because the distance measures calculated on such high dimensional data as 3D images have poor discrimination ability [45, 3]. On the other hand, our previously developed supervised dimension reduction approach [45] relies on the availability of training data in form of labeled subtomograms. Therefore, we employ the convolutional property of CNN to perform unsupervised feature extraction and dimension reduction. As small subvolumes are projected to a lower dimension, the distance between vectors is significantly more discriminative and faster to compute.

In principle, besides convolutional autoencoder, other alternative encoding approaches can also be employed, such as sparse coding [24], dictionary learning [1] and non-local means [8]. However, unlike convolutional autoencoder, these alternative approaches do not take the advantage of the inherent structures inside images such as local correlations and hierarchical organization of correlations. Also, some of these alternative approaches use linear representation models. Such approaches may fail to encode when the linearity assumption is invalid for certain data.

Data preparation and clustering processes still require the user to choose proper parameters such as the small subvolume size, the scale factor of Difference of Gaussian particle picking, and the number of k-means clusters. Currently we set an arbitrary number of 100 for clustering. We have tested Gap Statistic [40] and the Calinski-Harabasz index [30] for automatically choosing the cluster number. Both methods fail to converge to a certain cluster number. Since a simple manual grouping of resulting clusters is required, the impact of cluster number on the results is generally reduced. How to automatically determine the cluster numbers for encoded highly heterogeneous macromolecules from cellular tomograms remains as an open problem. Additionally, k-means clustering generates different cluster labels for each run, and thus the manual selection needs to be redone. A more user-friendly clustering procedure would be beneficial to further reduce the amount of manual work required.

Future works include 1) adapting the methods to take into account missing wedge effects; and 2) systematically optimize the metaparameters of the autoencoder and semantic segmentation models through various combinatorial optimization techniques [32], such as a line search, to further improve the performance of the models in terms of validation accuracy and training speed.

# 5 Software availability

Our approaches are distributed as open-source and can be downloaded for free by both academic and non-academic users from `http://cs.cmu.edu/~mxu1/software`.

# 6 Acknowledgement

We thank Dr. Robert Murphy for suggestions. We thank Dr. Zachary Freyberg and Kai Wen Wang for help with manuscript preparation. We thank Stephanie Siegmund for providing technical help. This work was supported in part by U.S. National Institutes of Health (NIH) grant P41 GM103712. X.Z and M.X acknowledges support of Samuel and Emma Winters Foundation. T.Z. acknowledges support of Sir Henry Dale Fellowship jointly funded by the Wellcome Trust and the Royal Society [Grant 107578/Z/15/Z] and Wellcome Trust Joint Infrastructure Fund Award 060208/Z/00/Z and Wellcome Trust Equipment Grant 093305/Z/10/Z to the Oxford Particle Imaging Centre.

## References

[1] Dictionary learning. *IEEE Signal Processing Magazine*, 28(2):27–38, 2011.

[2] Martín Abadi, Paul Barham, Jianmin Chen, Zhifeng Chen, Andy Davis, Jeffrey Dean, Matthieu Devin, Sanjay Ghemawat, Geoffrey Irving, Michael Isard, et al. Tensorflow: A system for large-scale machine learning. *arXiv preprint arXiv:1605.08695*, 2016.

[3] Charu C Aggarwal, Alexander Hinneburg, and Daniel A Keim. On the surprising behavior of distance metrics in high dimensional space. In *International Conference on Database Theory*, pages 420–434. Springer, 2001.

[4] Vijay Badrinarayanan, Alex Kendall, and Roberto Cipolla. Segnet: A deep convolutional encoder-decoder architecture for image segmentation. *arXiv preprint arXiv:1511.00561*, 2015.

[5] A. Bartesaghi, P. Sprechmann, J. Liu, G. Randall, G. Sapiro, and S. Subramaniam. Classification and 3D averaging with missing wedge correction in biological electron tomography. *Journal of structural biology*, 162(3):436–450, 2008.

[6] Alberto Bartesaghi, Guillermo Sapiro, and Sriram Subramaniam. An energy-based three-dimensional segmentation approach for the quantitative interpretation of electron tomograms. *IEEE Transactions on Image Processing*, 14(9):1314–1323, 2005.

[7] M. Beck, V. Lui, F. Förster, W. Baumeister, and O. Medalia. Snapshots of nuclear pore complexes in action captured by cryo-electron tomography. *Nature*, 449(7162):611–615, 2007.

[8] Priyam Chatterjee and Peyman Milanfar. A generalization of non-local means via kernel regression. In *Computational Imaging Vi*, page 68140, 2008.

[9] Muyuan Chen, Wei Dai, Ying Sun, Darius Jonasch, Cynthia Y He, Michael F Schmid, Wah Chiu, and Steven J Ludtke. Convolutional neural networks for automated annotation of cellular cryo-electron tomograms. *arXiv preprint arXiv:1701.05567*, 2017.

[10] Yuxiang Chen, Thomas Hrabe, Stefan Pfeffer, Olivier Pauly, Diana Mateus, Nassir Navab, and F Forster. Detection and identification of macromolecular complexes in cryo-electron tomograms using support vector machines. In *Biomedical Imaging (ISBI), 2012 9th IEEE International Symposium on*, pages 1373–1376. IEEE, 2012.

[11] Yuxiang Chen, Stefan Pfeffer, José Jesús Fernández, Carlos Oscar S Sorzano, and Friedrich Förster. Autofocused 3d classification of cryoelectron subtomograms. *Structure*, 22(10):1528–1537, 2014.

[12] François Chollet. keras. https://github.com/fchollet/keras, 2015.

[13] Javier Collado and Rubn Fernndez-Busnadiego. Deciphering the molecular architecture of membrane contact sites by cryo-electron tomography . *Biochimica et Biophysica Acta (BBA) - Molecular Cell Research*, 2017.

[14] Lidia Delgado, Gema Martínez, Carmen López-Iglesias, and Elena Mercadé. Cryo-electron tomography of plunge-frozen whole bacteria and vitreous sections to analyze the recently described bacterial cytoplasmic structure, the stack. *Journal of structural biology*, 189(3):220–229, 2015.

[15] Zachary Frazier, Min Xu, and Frank Alber. Tomominer and tomominercloud: A software platform for large-scale subtomogram structural analysis. *Structure*, 25(6):951–961, 2017.

[16] Ian Goodfellow, Yoshua Bengio, and Aaron Courville. *Deep Learning*. MIT Press, 2016. http://www.deeplearningbook.org.

[17] Kay Grünewald, Prashant Desai, Dennis C Winkler, J Bernard Heymann, David M Belnap, Wolfgang Baumeister, and Alasdair C Steven. Three-dimensional structure of herpes simplex virus from cryo-electron tomography. *Science*, 302(5649):1396–1398, 2003.

[18] Marion Jasnin, Mary Ecke, Wolfgang Baumeister, and Günther Gerisch. Actin organization in cells responding to a perforated surface, revealed by live imaging and cryo-electron tomography. *Structure*, 24(7):1031–1043, 2016.

[19] Simon Kemmerling, Stefan A Arnold, Benjamin A Bircher, Nora Sauter, Carlos Escobedo, Gregor Dernick, Andreas Hierlemann, Henning Stahlberg, and Thomas Braun. Single-cell lysis for visual analysis by electron microscopy. *Journal of structural biology*, 183(3):467–473, 2013.

[20] Diederik Kingma and Jimmy Ba. Adam: A method for stochastic optimization. *arXiv preprint arXiv:1412.6980*, 2014.

[21] Alex Krizhevsky, Ilya Sutskever, and Geoffrey E Hinton. Imagenet classification with deep convolutional neural networks. In *Advances in neural information processing systems*, pages 1097–1105, 2012.

[22] Yann LeCun, Yoshua Bengio, and Geoffrey Hinton. Deep learning. *Nature*, 521(7553):436–444, 2015.

[23] Yann LeCun, Léon Bottou, Yoshua Bengio, and Patrick Haffner. Gradient-based learning applied to document recognition. *Proceedings of the IEEE*, 86(11):2278–2324, 1998.

[24] Honglak Lee, Alexis Battle, Rajat Raina, and Andrew Y Ng. Efficient sparse coding algorithms. In *International Conference on Neural Information Processing Systems*, pages 801–808, 2006.

[25] Honglak Lee, Alexis Battle, Rajat Raina, and Andrew Y Ng. Efficient sparse coding algorithms. In *Advances in neural information processing systems*, pages 801–808, 2007.

[26] Jonathan Long, Evan Shelhamer, and Trevor Darrell. Fully convolutional networks for semantic segmentation. In *Proceedings of the IEEE Conference on Computer Vision and Pattern Recognition*, pages 3431–3440, 2015.

[27] Imanol Luengo, Michele C Darrow, Matthew C Spink, Ying Sun, Wei Dai, Cynthia Y He, Wah Chiu, Tony Pridmore, Alun W Ashton, Elizabeth MH Duke, et al. Survos: Super-region volume segmentation workbench. *Journal of Structural Biology*, 198(1):43–53, 2017.

[28] A. Martinez-Sanchez, I. Garcia, and J.J. Fernandez. A differential structure approach to membrane segmentation in electron tomography. *Journal of Structural Biology*, 175(3):372–383, 2011.

[29] Antonio Martinez-Sanchez, Inmaculada Garcia, and Jose-Jesus Fernandez. A ridge-based framework for segmentation of 3d electron microscopy datasets. *Journal of structural biology*, 181(1):61–70, 2013.

[30] Ujjwal Maulik and Sanghamitra Bandyopadhyay. *Performance Evaluation of Some Clustering Algorithms and Validity Indices*. IEEE Computer Society, 2002.

[31] Andrew Ng. Sparse autoencoder. *CS294A Lecture notes*, 72(2011):1–19, 2011.

[32] Jiquan Ngiam, Adam Coates, Ahbik Lahiri, Bobby Prochnow, Quoc V Le, and Andrew Y Ng. On optimization methods for deep learning. In *Proceedings of the 28th international conference on machine learning (ICML-11)*, pages 265–272, 2011.

[33] Fabian Pedregosa, Gaël Varoquaux, Alexandre Gramfort, Vincent Michel, Bertrand Thirion, Olivier Grisel, Mathieu Blondel, Peter Prettenhofer, Ron Weiss, Vincent Dubourg, et al. Scikit-learn: Machine learning in python. *The Journal of Machine Learning Research*, 12:2825–2830, 2011.

[34] Long Pei, Min Xu, Zachary Frazier, and Frank Alber. Simulating cryo electron tomograms of crowded cell cytoplasm for assessment of automated particle picking. *BMC bioinformatics*, 17(1):405, 2016.

[35] E.F. Pettersen, T.D. Goddard, C.C. Huang, G.S. Couch, D.M. Greenblatt, E.C. Meng, and T.E. Ferrin. UCSF Chimeraa visualization system for exploratory research and analysis. *Journal of computational chemistry*, 25(13):1605–1612, 2004.

[36] Prabhu Ramachandran and Gal Varoquaux. Mayavi: 3d visualization of scientific data. *Computing in Science & Engineering*, 13(2):40–51, 2011.

[37] A. Rigort, D. Günther, R. Hegerl, D. Baum, B. Weber, S. Prohaska, O. Medalia, W. Baumeister, and H.C. Hege. Automated segmentation of electron tomograms for a quantitative description of actin filament networks. *Journal of Structural Biology*, 2011.

[38] S.H.W. Scheres, R. Melero, M. Valle, and J.M. Carazo. Averaging of electron subtomograms and random conical tilt reconstructions through likelihood optimization. *Structure*, 17(12):1563–1572, 2009.

[39] Guang Tang, Liwei Peng, Philip R Baldwin, Deepinder S Mann, Wen Jiang, Ian Rees, and Steven J Ludtke. Eman2: an extensible image processing suite for electron microscopy. *Journal of structural biology*, 157(1):38–46, 2007.

[40] Robert Tibshirani, Guenther Walther, and Trevor Hastie. Estimating the number of clusters in a data set via the gap statistic. *Journal of the Royal Statistical Society*, 63(2):411–423, 2001.

[41] Svante Wold, Kim Esbensen, and Paul Geladi. Principal component analysis. *Chemometrics and intelligent laboratory systems*, 2(1-3):37–52, 1987.

[42] M. Xu, M. Beck, and F. Alber. High-throughput subtomogram alignment and classification by Fourier space constrained fast volumetric matching. *Journal of Structural Biology*, 178(2):152–164, 2012.

[43] M. Xu, S. Zhang, and F. Alber. 3d rotation invariant features for the characterization of molecular density maps. In *2009 IEEE International Conference on Bioinformatics and Biomedicine*, pages 74–78. IEEE, 2009.

[44] Min Xu, Martin Beck, and Frank Alber. Template-free detection of macromolecular complexes in cryo electron tomograms. *Bioinformatics*, 27(13):i69–i76, 2011.

[45] Min Xu, Xiaoqi Chai, Hariank Muthakana, Xiaodan Liang, Ge Yang, Tzviya Zeev-Ben-Mordehai, and Eric Xing. Deep learning based subdivision approach for large scale macromolecules structure recovery from electron cryo tomograms. *Bioinformatics*, 33(14):i13–i22, 2017.

[46] Min Xu, Elitza I Tocheva, Yi-Wei Chang, Grant J Jensen, and Frank Alber. De novo visual proteomics in single cells through pattern mining. *arXiv preprint arXiv:1512.09347*, 2015.

[47] Xiao Ping Xu, Christopher Page, and Niels Volkmann. *Efficient Extraction of Macromolecular Complexes from Electron Tomograms Based on Reduced Representation Templates*. Springer International Publishing, 2015.

# A convolutional autoencoder approach for mining features in cellular electron cryo-tomograms and weakly supervised coarse segmentation

Xiangrui Zeng[1], Miguel Ricardo Leung[2,3], Tzviya Zeev-Ben-Mordehai[2,3], and Min Xu*[1]

[1]Computational Biology Department, School of Computer Science, Carnegie Mellon University, Pittsburgh, 15213, USA
[2]Division of Structural Biology, Wellcome Trust Centre for Human Genetics, University of Oxford, Oxford, OX3 7BN, UK
[3]Cryo-electron Microscopy, Bijvoet Center for Biomolecular Research, Utrecht University, Utrecht, Netherlands

## Supplementary Document

## S1 Details of convolutional neural network

The learnable unit of each layer, called a neuron, connects with part of a previous layer, called a receptive field or filter. There exist different types of CNN layers: convolutional layer, fully connected layer, up sampling layer, pooling layer, etc. The input and output size of a CNN layer can be arbitrary and defined by tuning the specific parameters of that layer. For convolutional layers, filter (a.k.a kernel) refers to the receptive field that convolves the input. Stride refers to the shifting size of each convolving step. A convolution layer can be seen as set of learned feature extractors [1]. Given $N$ filters of size $K^3$ and stride S, the output y is processed by sliding each filter with steps of $S$ on the input. Summing up the dot product at every location will give the output value y.

Let the input to the layer be $x$ with $Z$ filter maps, the filters of the layer be $W$ and the output be $y$. Mathematically [7], the convolutional layer is defined by

$$y^m_{i,j,k} = \sum_{p=0}^{Z-1}\sum_{a=0}^{K-1}\sum_{b=0}^{K-1}\sum_{c=0}^{K-1} W^m_{a,b,c} \cdot x^p_{a+S\cdot i,b+S\cdot j,c+S\cdot k} \tag{1}$$

, where, $y^m_{i,j,k}$ is the index $(i,j,k)$ of the $m^{\text{th}}$ output volumetric slice , $W^m_{a,b,c}$ is the index $(a,b,c)$ of the $m^{\text{th}}$ filter, and $X^p_{a+S\cdot i,b+S\cdot j,c+S\cdot k}$ is the index $(a+S\cdot i, b+S\cdot j, c+S\cdot k)$ of the $p^{\text{th}}$ input filter map.

A pooling layer is a type of non-linear down sampling layer to reduce the spatial representation size. Such pooling often outputs the local maximum (max pooling) or average (average pooling) of its receptive field. Mathematically, the max pooling layer is represented by

$$y^m_{i,j,k} = \max_{a,b,c\in[0,K-1]}(x^p_{a+S\cdot i,b+S\cdot j,c+S\cdot k}) \tag{2}$$

.

Similarly, an up sampling layer is to increase the spatial representation size by repeating the number in the receptive field. Mathematically, the up sampling layer is represented by

$$y^m_{a+S\cdot i,b+S\cdot j,c+S\cdot k} = x^p_{i,j,k} \qquad \forall a,b,c \in [0,K-1] \tag{3}$$

*Corresponding author email: mxu1@cs.cmu.edu

.

A flatten layer is a layer that flattens the input to a vector. For example, inputting array of size $3 \times 3 \times 3$, a flatten layer will output a vector of length 27, without breaking the order of the input. Mathematically, the flatten layer is represented by

$$y_{M \cdot N \cdot k + M \cdot j + i} = x_{i,j,k} \tag{4}$$

, where the input array is of size $M \times N \times L$.

Following a flatten layer, a fully connected layer can be applied. A fully connected layer is a layer in which every neuron is connected to every neuron in the previous layer. The input and output of a fully connected layer can both be arbitrary dimension of arbitrary numbers of neurons. Mathematically, a fully connected layer of one-dimensional input and output is defined as

$$y_i = \sum_{a=1}^{L} W_a^i \cdot x_a \tag{5}$$

, where the input is a vector of length $L$ and $W_a^i$ is the weight connecting input $x_a$ and output $y_i$.

For our Autoencoder3D model, following a flatten layer, a fully connected layer of 32 neurons will output a vector of length 32 as the encoding for a small subvolume. We use the stack of convolution operations in CNN to take advantage of the inherent structure in the small subvolumes, i.e. the local correlations and hierarchical organization of correlations. Therefore, our 3D convolutional autoencoder can encode a complex 3D small subvolume into a simple vector, so that the clustering can be directly and efficiently computed using such encoded vectors. Flatten and fully connected layers are necessary for constructing such encoded vector. In order to train the encoder, it is necessary to decode the encoded vector into a decoded 3D small subvolume. The first step of such decoding requires using a fully connected layer to compute a vector $y$ of length $M \cdot N \cdot L$ from the encoded vector. Then the unflattening step is to reshape $y$ into a 3D array of size $M \times N \times L$. Such array is then used further reconstruction through upsampling and convolution operations.

Following a convolution, an activation function such as a rectified liner unit (ReLU) [5] is applied. The ReLU activation is defined as $o^{ReLU}(x) = \max\{0, x\}$. ReLU activation has the advantage of being sparse, scale invariant, and simple enough for efficient computation and gradient propagation.

For an autoencoder network, the output reconstructs the input. Because both the input and output consist continuous real values, the final output layer is usually a linear activation layer. The linear activation is defined as $o^{linear}(x) = x$, which is simple to compute and favored when the output is composed of unbounded continuous values. Semantic segmentation is one type of multi-class classification tasks where an image is partitioned into semantically meaningful parts and each part is classified into one of the pre-defined classes. Softmax activation is usually applied as the last output layer for classification tasks. The softmax activation is defined as:

$$o_j^{softmax}(x) = P(j|x) = \frac{e^{x^T w_j}}{\sum_{l=1}^{L} e^{x^T w_j}}$$

where $x$ is the input of the last layer and $w_j$ is the weight associated with the $j$th class. The softmax activation calculates the probability $P(j|x)$ of a sample being classified into each class. In the 3D image semantic segmentation task, each voxel is treated as a sample and a semantic class label is assigned to each voxel.

The Adam optimizer uses adaptive gradient that updates the learning rate to facilitate convergence. In the paper [4] that introduced Adam, $\beta_1 = 0.9$ and $\beta_2 = 0.99$ were recommended as the parameters. Since then, the widespread success with the Adam optimizer using these parameters has made them the default parameters for Adam. For this reason and for our success with these parameters, we used $\beta_1 = 0.9$ and $\beta_2 = 0.99$.

Categorical cross-entropy is a loss function, which is usually coupled with softmax final layer for multi-class classification task. Formally, the cross-entropy loss is defined as

$$L_j = -\sum_j x_j \log \hat{x_j} \tag{6}$$

, where $x_j$ denotes the true probability of $x$ belonging to class $j$ and $\hat{x_j}$ denotes the predicted probability of $x$ belonging to class $j$. From a probabilistic point of view, minimizing the cross-entropy loss can be seen as minimizing negative log likelihood of the correct label, which is similar to performing a maximum likelihood estimation. Cross-entropy is the loss that corresponds to the Softmax activation function, and has become default for multi-class classification problems that use Softmax for activation.

A CNN model is trained to optimize the weights of each neuron in each layer through backpropagation in conjugation with an optimization method, such as gradient descent. After being provided with training data of paired inputs and outputs, the training works by forward-propagating the input through the network and comparing the error of output, which is then back-propagated to calculate the gradient of the loss function with respect to each neuron's weight. The weights are then updated according to the gradient by gradient descent. Gradient descent methods are the most common optimization methods [3] because of their computational simplicity.

# S2 Gaussian smoothing

The Gaussian smoothing (a.k.a Gaussian blur) is an imaging processing technique that blurs an image by applying a Gaussian function. Typically, after Gaussian smoothing, the noise and detail of the image are both reduced. For each voxel, Gaussian smoothing replaces its value by a weighted average of its neighboring voxels by the Gaussian function. The 3D Gaussian function is defined as

$$G(x,y,z) = \frac{1}{(2\pi\delta^2)^{\frac{3}{2}}} e^{\frac{x^2+y^2+z^2}{2\delta^2}} \tag{7}$$

, where $x, y, z$ denotes the distance from the current voxel and $\delta$ denotes the standard deviation. Gaussian function calculates the weights of neighboring voxels and uses the weight average as the new value of the voxel. During Gaussian smoothing, the Gaussian function is applied to every voxel and replaces its value by the weighted average of its neighboring voxels. $\delta$ (in the unit of number of voxels) is a parameter that controls the magnitude of the blur. When $\delta$ is high, more weight is placed on the neighboring voxels and the image is blurred to a greater extent. In Figure S1, We show examples of applying Gaussian smooth with different $\delta$ to a surface structure small subvolume from COS-7 tomogram 1. Clearly, the noise is reduced when we apply a Gaussian smoothing with $\delta = 2.0$. However, when $\delta = 5.0$, the image becomes too blurry with a large loss of structural information of interest.

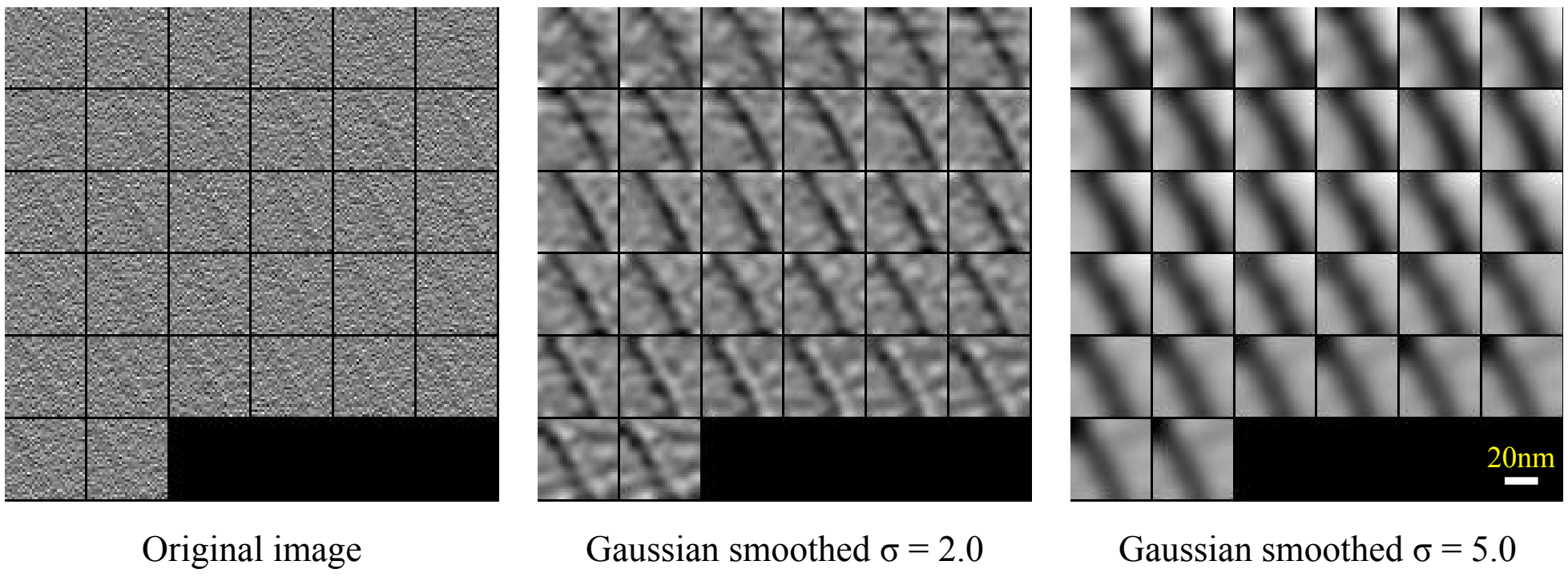

Figure S1: Left: original 2D slices of a small subvolume containing a surface structure. Middle: Gaussian smoothed small subvolume with $\delta = 2.0$. Right: Gaussian smoothed small subvolume with $\delta = 5.0$.

In our autoencoder based image feature classification process, Gaussian smoothing is a useful preprocessing step to compensate the variation of defocus and choices of CTF correction. A tomogram with a small defocus would have a lower contrast. In this paper we only use our autoencoder and semantic segmentation models for low resolution separation and segmentation of image features instead of high resolution recovery of fine structural details of macromolecular complexes. Therefore we used Gaussian smoothing preprocessing step to enhance the contrast of tomograms. The level of Gaussian smoothing is properly chosen so that the image features of ultrastructures are clearly visualized. Such Gaussian smoothing compensates variations resulted from different defocus and CTF correction choices. Therefore, the performance of our low resolution image feature separation and segmentation is expected to be unaffected by the defocus and CTF correction.

## S3 Numerical study

In order to quantitatively assess the performance of our autoencoder model, we performed a numerical study on a realistically simulated tomogram to analyze the accuracy of our model in terms of precision and recall. Precision is defined as

$$Precision = \frac{number\ of\ true\ positives}{number\ of\ true\ positives\ +\ number\ of\ false\ positives} \tag{8}$$

, which measures the percentage of small subvolumes containing true structures among all selected small subvolumes.

Recall is defined as

$$Recall = \frac{number\ of\ true\ positives}{number\ of\ true\ positives\ +\ number\ of\ false\ negatives} \tag{9}$$

, which measures the percentage of selected small subvolumes containing true structures among all small subvolumes containing the true structure. The missing wedge effect is also simulated and discussed in the numerical study.

We simulated a tomogram according to the experimental conditions of COS-7 tomogram 1 ('tilt angle range' : $\pm 60°$, 'tilt angle increment': 4°, 'voxel size': 1.42 $nm^3$ , 'defocus': -6 $\mu$m, 'voltage': 300V, 'spherical aberration': 2.0 $mm$).

We aligned the blob-like small subvolumes in COS-7 tomogram 1 against the Ribosome template and calculated the Pearson correlations of randomly sampled pairs of subtomograms among the top 100 hits. We estimated the SNR distribution using such Pearson correlations according to [2]. The mean SNR was 0.0044. We adjust the simulated tomograms so that the extracted Ribosome small subvolumes also have similar SNR to the SNR experimental small subvolumes calculated above.

We added a small empty sphere of inner radius 63 voxels and outer radius 70 voxels in the center to represent sub-cellular vesicles, half of a big empty sphere of inner radius 203 voxels and outer radius 210 in the right to represent part of a cell. The thicknesses of both spheres are about 10$nm$, which is similar to the thickness of a cellular membrane. We also added 8 planes on the right to represent surface structures due to imaging artifact such as tomogram boundaries. 776 human ribosome structures (PDB ID: 5T2C) are randomly rotated and added to the simulated tomogram with equal distance between them and separable from surface structures. The final simulated tomogram has a size of $400 \times 400 \times 1200$ voxels. One XZ plane of the simulation ground truth (electron density maps) is plotted in Figure S2 (a).

To eliminate the influence of particle picking on the numerical analysis, we selected small subvolumes from their true locations rather than using the Difference of Gaussian particle picking, as we did for real tomograms. 776 ribosome structure small subvolumes and 1535 surface structure small subvolumes were selected. We also selected 2311 null small subvolumes from the non-structural region. All 4622 selected small subvolumes are of size $32^3$, the same dimension as real tomograms. Since the simulated tomogram has mean of 0, the negative regions generally indicate signal regions. We pose normalized these small subvolumes only based on the signal regions and trained an Autoencoder3D model to cluster homogeneous structures. For real tomograms where the threshold for signal regions are unknown, we recommend using the value normalization procedure described in the Section 3.2. Because the total number of small subvolumes is smaller than in real tomograms, we applied k-means clustering algorithm to cluster the encoded small subvolumes into 30 clusters.

During the data acquisition procedure, since the sample cannot be imaged in full 180°tilt angle range, the missing angular range will result in poor data quality in some regions of the reconstructed tomogram, which is known as missing wedge effect [6]. When we simulated a noise-free tomogram based on the electron density map, we can clearly observe the missing wedge effect. In Figure S2 (b), the two red arrows indicate part of the vertical surfaces whose signal is weakened due to the missing wedge effect.

In Figure S2 (d), the two green arrows indicate the weakened vertical surface region due to the missing wedge effect. Those weakened regions were still successfully clustered into surface structural clusters and embedded into the tomogram by our Autoencoder3D network. We found that when we used the Difference of Gaussian particle picking method, the weakened surface regions were not picked up as small subvolumes. Therefore, provided that the small subvolumes of regions affected by missing wedge be picked out, to certain extent, our Autoencoder3D network was able to detect weakened signals in small subvolumes due to missing wedge effect. However, improving the particle picking method to successfully pick up small subvolumes of regions affected by missing wedge is out of the scope of this paper. We note here that our method is tested on dataset with tilt angle range $\pm 60^\circ$. The impact of missing wedge effect is subject to further inspection especially for the tomograms with tilt angle range smaller than $\pm 60^\circ$.

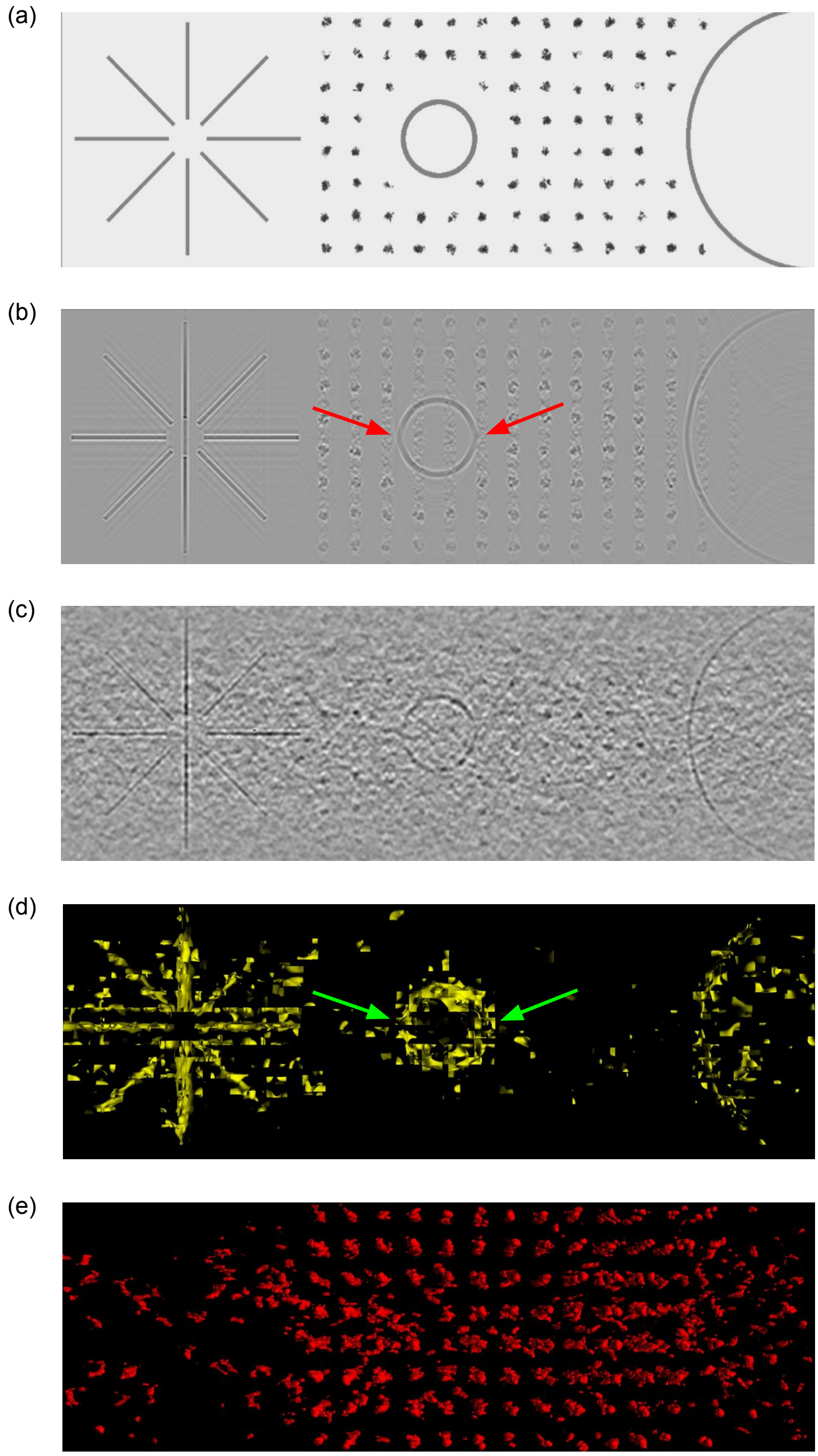

Figure S2: The representative XZ plane of (a) ground truth (electron density map) of simulated tomogram; (b) an example of a simulated tomogram with no noise (red arrows indicate major missing wedge effects); (c) realistically simulated tomogram according to COS-7 tomogram 1; (d) isosurface of selected surface structural small subvolumes embedded in the simulated tomogram; (e) isosurface of selected ribosome small subvolumes embedded in the simulated tomogram. The X axis is in the vertical direction, the Z axis is in the horizontal direction.

In Figure S2 (d), we selected 4 clusters as ribosome structural clusters for annotation, which resulted in a precision of 66.02% and recall of 61.34% for ribosome structures. We selected 14 clusters as surface structural clusters for annotation, which resulted in a precision of 73.92% and recall of 44.69% for surface structures. We note here that since manual selection is required, selecting more clustering will result in a higher precision but lower recall. We plotted the precision and recall curves for ribosome and surface structures in Figure S3. When more clusters are selected, precision decreases but recall increases. A moderate selection will result in precision and recall higher than 60% for both ribosome and surface structures, which is considered to be accurate for a tomogram with such low SNR.

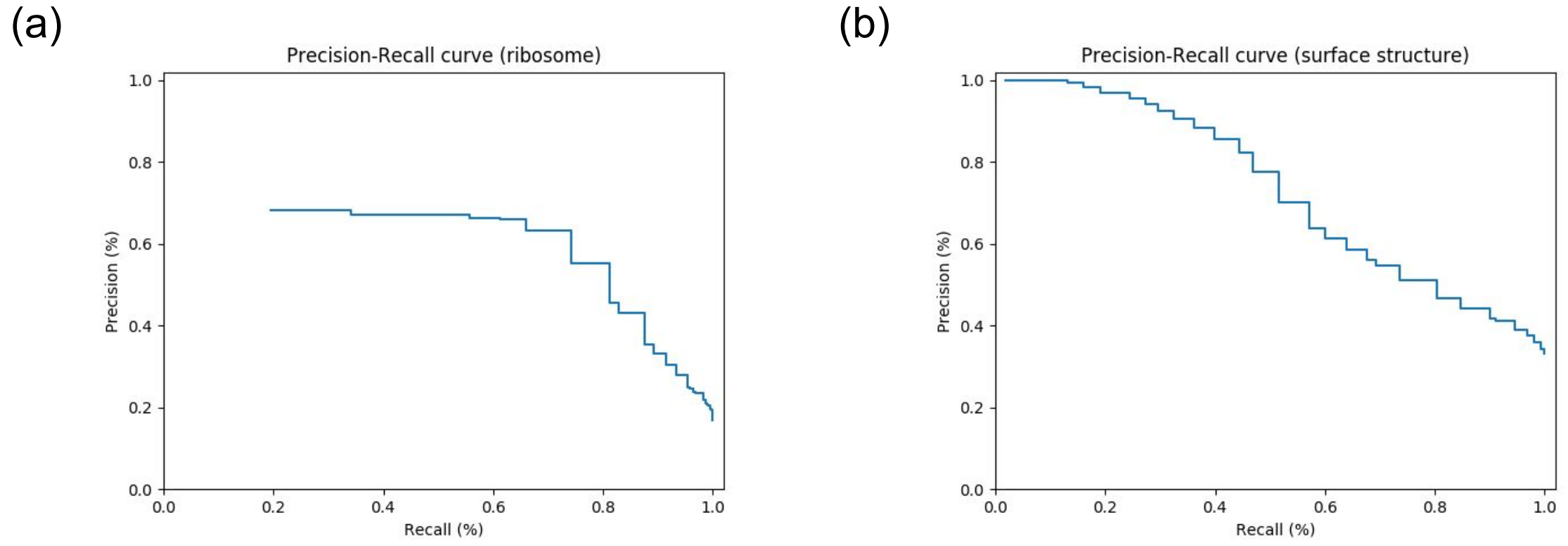

Figure S3: (a) precision-recall curve of selecting ribosome structures in numerical study. (b) precision-recall curve of selecting surface structures in the numerical study.

# S4 Inspection and averaging of ‘large globule’ small subvolumes

**Template search**

We selected 50 representative macromolecular complexes (Table S1, including 2 ribosome structures) and generated 50 structural templates according to the imaging parameters of the COS-7 tomograms. All 384 selected large globule small subvolumes of 9 clusters were aligned against each of the 50 structural templates (denoted by PDB ID) using our previously developed fast alignment method [8]. Figure S4 shows a box plot of the distribution of alignment scores (Pearson correlation of template and aligned small subvolume) of the large globule small subvolumes of against 50 different template complexes. The two ribosome structural templates (bacterial ribosome (PDB ID: 2AWB) and human ribosome (PDB ID: 5T2C)) resulted in the highest alignment score distributions.

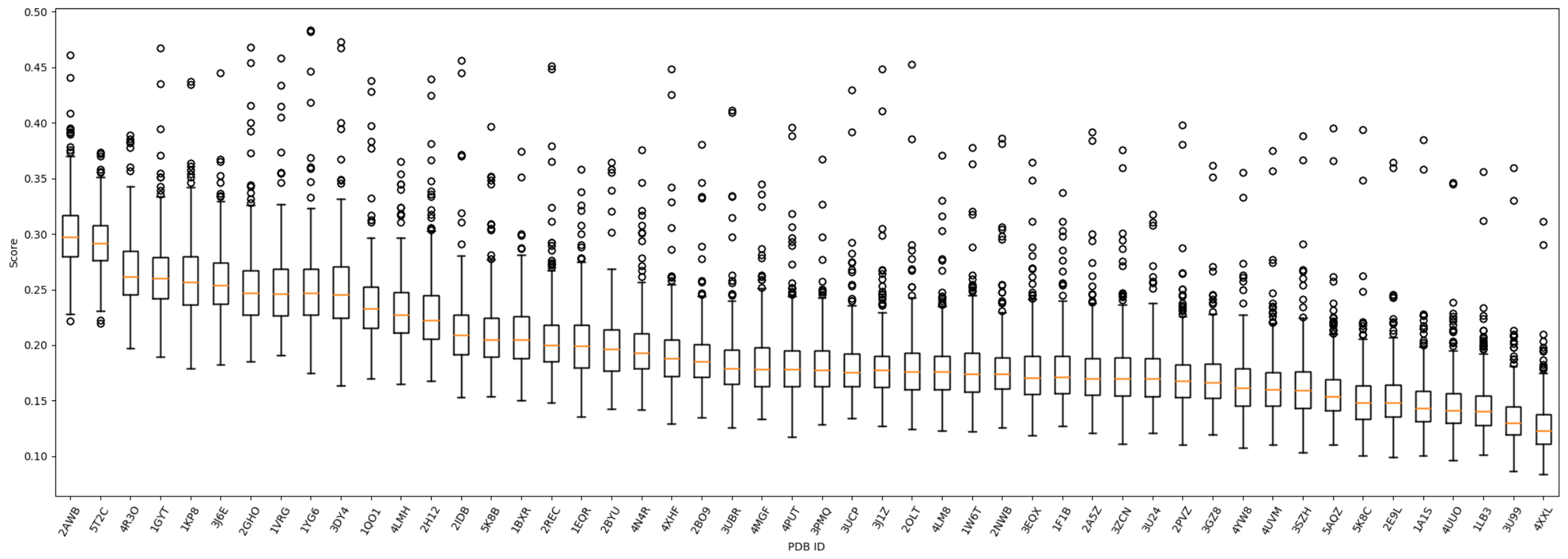

Figure S4: Box plot of the distribution of alignment scores of the large globule small subvolumes of against 50 different template complexes.

| PDB ID | Macromolecular Complex |
|---|---|
| 2AWB | Bacterial ribosome |
| 5T2C | Human ribosome |
| 4R3O | Human 20S Proteasome |
| 1GYT | E. coli Aminopeptidase A |
| 1KPB | PKCI-1-APO |
| 3J6E | Microtubules stabilized by GmpCpp |
| 2GHO | Thermus aquaticus RNA polymerase |
| 1VRG | Propionyl-CoA carboxylase |
| 1YG6 | ClpP |
| 3DY4 | Yeast 20S proteasome |
| 1QO1 | Rotary Motor in ATP Synthase |
| 4LMH | Outer membrane decaheme cytochrome OmcA |
| 2H12 | Acetobacter aceti citrate synthase |
| 2IDB | 3-octaprenyl-4-hydroxybenzoate decarboxylase |
| 5K88 | Tris-thiolate Pb(II) Complex |
| 1BXR | Carbamoyl phosphate synthetase |
| 2REC | RECA hexamer |
| 1EQR | Aspartyl-TRNA synthetase |
| 2BYU | M.tuberculosis Acr1(Hsp 16.3) |
| 4N4R | Structure basis of lipopolysaccharide biogenesis |
| 4XHF | Shewanella oneidensis NqrC |
| 2BO9 | Human carboxypeptidase A4 |
| 3UBR | Shewanella oneidensis cytochrome-c Nitrite Reductase |
| 4MGF | Apo-PhuS |
| 4PUT | Arabidopsis thaliana TOP2 oligopeptidase |
| 3PMQ | Outer membrane decaheme cytochrome MtrF |
| 3UCP | Outer membrane Endecaheme cytochrome UndA |
| 3J1Z | Zinc Transporter YiiP |
| 2OLT | Phou-like protein |
| 4LM8 | Outer membrane decaheme cytochrome MtrC |
| 1W6T | Octameric Enolase |
| 2NWB | Putative 2,3-dioxygenase |
| 3EQX | A fic family protein |
| 1F1B | E. coli asparate transcarbamoylase P268A |
| 2A5Z | Shewanella oneidensis MR-1 |
| 3ZCN | Fic protein with ATP |
| 3U24 | A putative lipoprotein |
| 2PVZ | Methylaconitate isomerase PrpF |
| 3GZ8 | NUDIX domain of Shewanella oneidensis NrtR |
| 4YW8 | Rat cytosolic pepck |
| 4UVM | POT family transporter PepTSo |
| 3SZH | Apo shwanavidin |
| 5AQZ | HSP72 with adenosine-derived inhibitor |
| 5K8C | 3-deoxy-alpha-D-manno-octulosonate 8-oxidase |
| 2E9L | Human Cytosolic Neutral beta-Glycosylceramidase |
| 1A1S | Ornithine carbamoyltransferase |
| 4UUO | Apo Trichomonas vaginalis malate dehydrogenase |
| 1LB3 | Mouse L chain ferritin |
| 3U99 | Diheme cytochrome type c |
| 4XXL | Class 1 cytochrome MtoD |

Table S1: The experimental macromolecular complexes used for template search.

**Reference-free averaging**

Of the 9 'large globule' image feature clusters, we selected 5 whose decoded cluster center have diameters that are most similar to the diameter of a ribosome. This selection was to reduce the number of false-positives in the selected clusters. Then, we performed a reference-free alignment of all 184 small subvolumes in the 5 clusters. We note here that aligning and averaging is difficult due to several reasons: 1) the number of small subvolumes (subtomograms) is small; 2) the small subvolumes may likely contain very heterogeneous structures because they are obtained from coarse classification; 3) the SNR of the small subvolumes is low (0.0044); and 4) tilt series was captured with 4° interval. We first averaged all the small subvolumes. Then for each iteration, we aligned each small subvolume with the average. The top 50% small subvolumes with highest alignment scores were used to generate the new average. After 50 iterations, the average has already converged. Figure S5 shows that the average has similar size as compared to the ribosome template.

(a) (b)

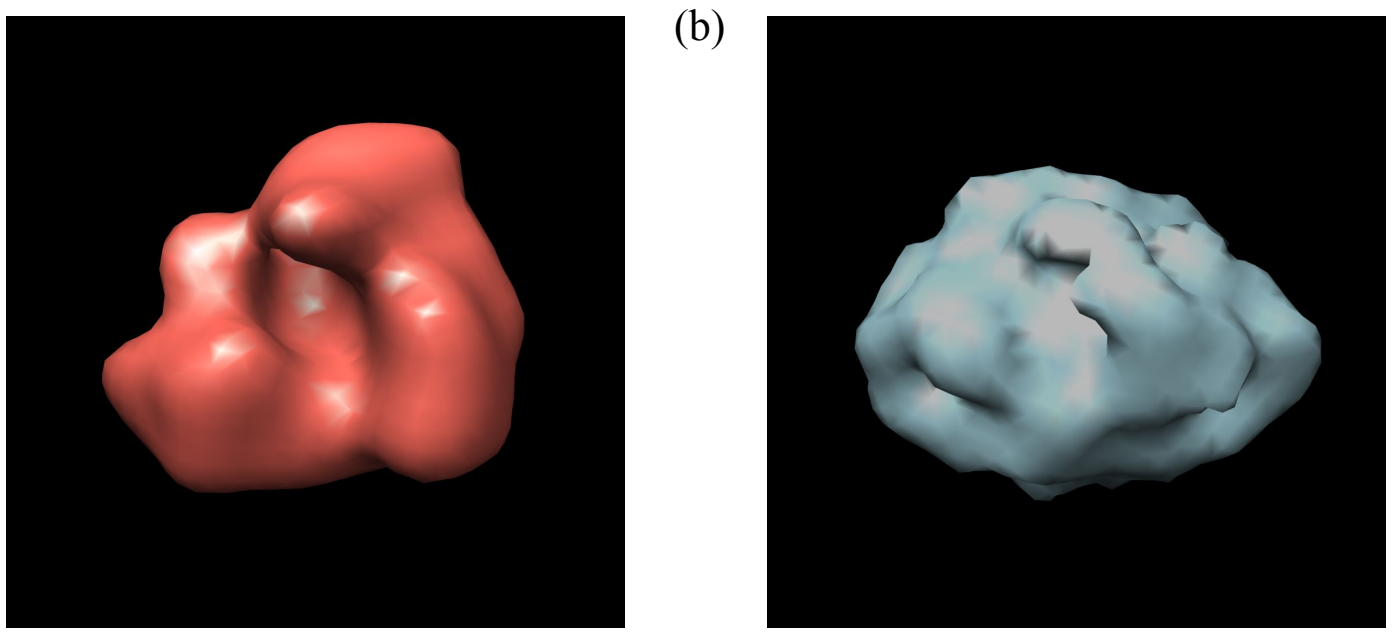

Figure S5: (a) Isosurface of ribosome template filtered at 80 Å (PDB ID: 5T2C). (b) Isosurface of averaged large globule small subvolumes

# S5 Computing time details

| Step | Number of inputs | Input size | Number of epoches | Total time | Environment |
|---|---|---|---|---|---|
| Particle picking | 1 | $959 \times 928 \times 293$ | NA | 15 min | scipy 0.19.1, numpy 1.13.1 |
| Pose normalization | 38112 | $32^3$ | NA | 20 min | numpy 1.13.1 |
| Autoencoder3D training | 34300 | $32^3$ | $\sim$80 | 10 hours | Keras 2.0.8 |
| Encoder3D prediction | 38112 | $32^3$ | NA | 45 s | Keras 2.0.8 |
| K-means clustering | 38112 | 32 | NA | 3 min | scikit-learn 0.18.1 |
| Manual selection | 100 | $32^3$ | NA | $\sim$ 10 - 30 min | |
| EDSS3D training | 727 | $32^3$ | $\sim$100 | 13 min | Keras 2.0.8 |
| EDSS3D prediction | 312 | $32^3$ | NA | 3 s | Keras 2.0.8 |

Table S2: Details of expected durations (including possible overheads) for different steps in sequence. All steps were tested under the environment of Python 2.7 in Ubuntu 16.04. Keras 2.0.8 was backended by tensorflow-gpu 1.0.1 and has the dependencies CUDA 8.0.61 and cuDNN v5.

## References

[1] Ben Athiwaratkun and Keegan Kang. Feature representation in convolutional neural networks. *arXiv preprint arXiv:1507.02313*, 2015.

[2] Joachim Frank. *Three-dimensional electron microscopy of macromolecular assemblies: visualization of biological molecules in their native state*. Oxford University Press, 2006.

[3] Ian Goodfellow, Yoshua Bengio, and Aaron Courville. *Deep Learning*. MIT Press, 2016. http://www.deeplearningbook.org.

[4] Diederik Kingma and Jimmy Ba. Adam: A method for stochastic optimization. *arXiv preprint arXiv:1412.6980*, 2014.

[5] Vinod Nair and Geoffrey E Hinton. Rectified linear units improve restricted boltzmann machines. In *Proceedings of the 27th international conference on machine learning (ICML-10)*, pages 807–814, 2010.

[6] Lassi Paavolainen, Erman Acar, Uygar Tuna, Sari Peltonen, Toshio Moriya, Pan Soonsawad, Varpu Marjomäki, R Holland Cheng, and Ulla Ruotsalainen. Compensation of missing wedge effects with sequential statistical reconstruction in electron tomography. *PloS one*, 9(10):e108978, 2014.

[7] Andrea Vedaldi and Karel Lenc. Matconvnet: Convolutional neural networks for matlab. In *Proceedings of the 23rd ACM international conference on Multimedia*, pages 689–692. ACM, 2015.

[8] M. Xu, M. Beck, and F. Alber. High-throughput subtomogram alignment and classification by Fourier space constrained fast volumetric matching. *Journal of Structural Biology*, 178(2):152–164, 2012.